# A family of air-stable chalcogenide solid electrolytes in $Li_2BMQ_4$ (B = Ca, Sr and Ba; M = Si, Ge and Sn; Q = O, S and Se) systems

Huican Mao[1, 2#], Xiang Zhu[2#], Guangmao Li[3#], Jie Pang[2], Junfeng Hao[2], Liqi Wang[2], Hailong Yu[2], Youguo Shi[2], Fan Wu[2*], Shilie Pan[3*], Ruijuan Xiao[2*], Hong Li[2*], Liquan Chen[2]

1. Department of Energy Storage Science and Engineering, School of Metallurgical and Ecological Engineering, University of Science and Technology Beijing, Beijing 100083, China

2. Beijing National Laboratory for Condensed Matter Physics, Institute of Physics, Chinese Academy of Sciences, 100190, Beijing, China.

3. Research Center for Crystal Materials, Xinjiang Technical Institute of Physics and Chemistry, Chinese Academy of Sciences, Urumqi 830011, China.

E-mail: fwu@iphy.ac.cn, slpan@ms.xjb.ac.cn, rjxiao@iphy.ac.cn, hli@iphy.ac.cn

# Abstract

Combining high-throughput first-principles calculations and experimental measurements, we have identified a novel family of fast lithium-ion chalcogenide conductors in $Li_2BMQ_4$ (2114, B = Ca, Sr and Ba; M = Si, Ge and Sn; Q = O, S and Se) systems. Our calculations demonstrate that most of the thermodynamically and kinetically stable sulfides and selenides in this new system exhibit ultralow $Li^+$ ion migration activation energy (0.16 eV ~ 0.56 eV) and considerable bandgaps varying between ~ 2 eV and 3.5 eV. We have successfully synthesized $Li_2BaSnS_4$ and $Li_2SrSiS_4$, and they exhibit excellent moisture stability through $H_2S$ gas measurements. Electrochemical impedance measurements indicate 2114 systems show the typical features of solid ionic conductors, with a room-temperature $Li^+$ conductivity close to $5\times10^{-4}$ mS/cm aligning with our molecular dynamics simulations. Furthermore, we have theoretically investigated the substitution of $Cl^-$ at $S^{2-}$ site. The doped compounds display significantly higher conductivity, with an increase of about three orders of magnitude (up to a maximum of 0.72 mS/cm) compared to the undoped compounds. These findings offer valuable insights for the further exploration of potential chalcogenide solid electrolyte materials with robust air stability and enhanced ionic conductivity for practical applications in lithium-ion batteries.

## Introduction

In the race to decarbonize the global economy, all-solid-state battery (ASSB) technology is emerging as one of the most promising solutions to address the energy/power density and safety problems associated with combustible liquid electrolytes in lithium-ion batteries (LIBs) [1-6]. Solid electrolytes (SEs) with fast ionic transport, outstanding deformability and improved air stability are critical components in the development of next-generation ASSBs [7-9]. Despite significant research efforts, only a limited number of inorganic materials out of tens of thousands of known inorganic materials for LIBs, such as oxide-based SEs ($Li_{1.3}Al_{0.3}Ti_{1.7}(PO_4)_3$[10], $Li_{6.55}La_3Zr_2Ga_{0.15}O_{12}$[11], $Li_{0.34}La_{0.51}TiO_{2.94}$ [12] etc.), sulfide-based SEs ($Li_{10}GeP_2S_{12}$[13], $Li_7P_3S_{11}$[14], $Li_3PS_4$ [15], $Li_6PS_5Cl$ [16], $Li_4SnS_4$ [17] etc.) and halide-based SEs ($Li_3InCl_6$ [18] etc.), have been identified as fast ionic conductors. However, these above fast ionic conductors still suffer from various challenges, such as inflexibility and expensive large-scale production of oxides, moisture sensitivity and poor compatibility with cathode materials for sulfides, and vulnerability to moisture and low oxidation voltage for halides [19]. Among them, air stability could be a common issue for oxide-, sulfide- and halide-based SEs [20-23]. For example, when oxide-based SEs are exposed to humid air, the sluggish exchange between $Li^+$ and $H^+$ could form LiOH and $Li_2CO_3$, increasing interfacial resistance [24, 25]. For Halide-based SEs, such as $Li_3InCl_6$, it is easy to absorb water, and first becomes a crystalline hydrate, then partially decomposes to $InCl_3$, LiCl and $In_2O_3$ [26]. Furthermore, except for the reduction of ionic conductivity and hydrolysis evolution, sulfide-based SEs are prone to release toxic gas $H_2S$ and result in structure/performance degradation [27-29], which severely limits the practical application of chalcogenide-based ASSBs.

Although sulfide-based SEs suffer from the most deadly and well-known drawback of poor air stability, they have been recognized as one of the most promising SEs due to their exceptional superionic conductivity and excellent deformability [13, 30, 31]. To enhance the air stability of sulfide-based SEs, a few promising strategies were employed including partial substitution of P/S by softer acids/O, and designing new materials with softer acids as central cations [20-23] based on the hard and soft acids and bases theory (HSAB). For example, Xu et al. synthesized $Li_{6.25}PS_4O_{1.25}Cl_{0.75}$ by partial oxygen substitution at both S and Cl sites, achieving high stability toward moist air [32]. The substitution of P with softer acid Sb in $Li_{10}GeP_2S_{12}$ electrolyte system leads to the formation of robust bond Sb-S, which exhibits an effective suppression for $H_2S$ gas generation [29]. In 2012, Kaib et al first synthesized a novel fast ionic conductor $Li_4SnS_4$, which shows good moisture stability [33].

While elemental substitution is an effective approach to improve air stability of sulfide SEs, partial substitution of P/S with softer acids/O may cause irreversible structural degradation or the reduction of ionic conductivity [29, 32]. Based on HSAB theory, complete substitution of P with softer acids (e.g., Sn, Sb, Ge) may provide the ideal configuration for enhanced significantly air stability. After decades of research, so far, only Li-Sn/Sb-S system demonstrating superior air stability has been discovered as the new materials of sulfide-based SEs with softer acids as central cations [28, 33, 34]. Hence, further investigations into novel compositions/structures of sulfide SEs with outstanding air stability and ionic conductivity are required to diversify the range of sulfides SEs and facilitate the development of next-generation high-performance, safe and sustainable sulfide-based ASSBs.

In this work, we combine materials database with bond valence (BV) and density functional theory (DFT) calculations to build up the high-throughput screening platform for searching for promising chalcogenide-based SEs. Based on this high-throughput calculations, we find a novel family of air-stable fast lithium-ion chalcogenide conductors in $Li_2BMQ_4$ (2114, B = Ca, Sr and Ba; M = Si, Ge and Sn; Q = O, S and Se) systems, which possess ultralow $Li^+$ migration energy barrier (0.16 ~ 0.56 eV) and considerable bandgap between ~2.0 and 3.5 eV for sulfides and selenides (**Table S2**). The experimentally synthesized $Li_2BaSnS_4$ and $Li_2SrSiS_4$ display exceptional air stability by total $H_2S$ experiment and a promising room-temperature ionic conductivity around $5\times10^{-4}$ mS/cm by impedance spectroscopy measurements, which aligns with our calculations. More importantly, the introduction of Li vacancy by substituting $S^{2-}$ with $Cl^-$ could enhance ionic conductivity by three orders of magnitude based on first-principles molecular dynamics (FPMD) simulations.

## Results and Discussion

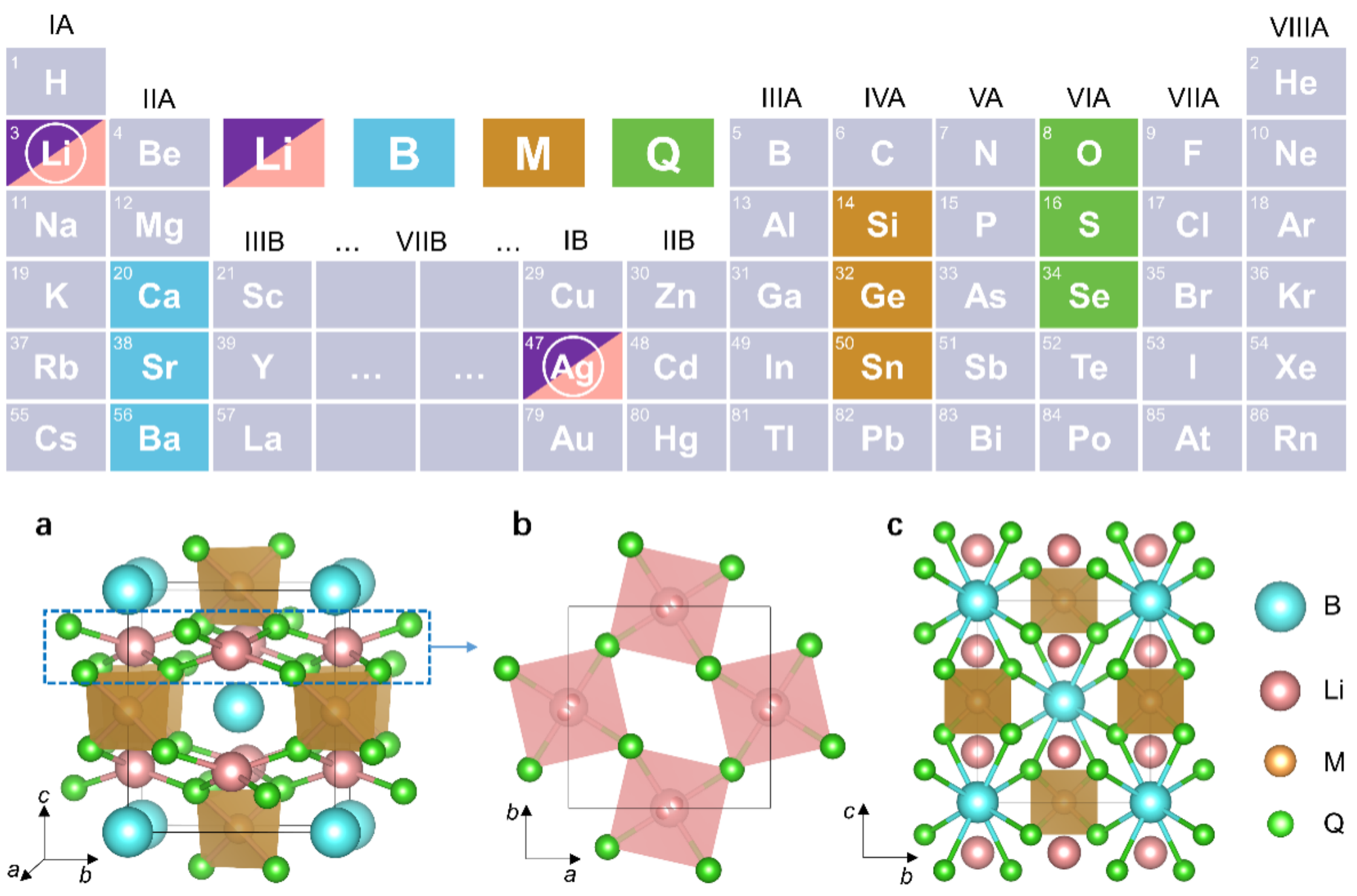

**Fig.1 The schematic diagram of crystal structures in $Li_2BMQ_4$ (B = Ca, Sr and Ba; M = Si, Ge and Sn; Q = O, S and Se) system. (a)** The crystal structure of $Li_2BMQ_4$ with a 2D Li-Q layer is plotted in the blue dashed rectangle. Some of $Li_2BMQ_4$ structures are obtained from the substitution of Li at Ag site in $Ag_2BMQ_4$ systems. **(b)** The 2D layer is composed of the linking $LiQ_4$ units by sharing corners. **(c)** The $BQ_8$ dodecahedra connects to each other by edge-sharing $MQ_4$/corner-sharing $BQ_8$ units.

## Crystal Structure

We first obtain the potential lithium conductors $Li_2BMQ_4$ (B = Ca, Sr and Ba; M = Si, Ge and Sn; Q = O, S and Se) systems (**Fig.1**) by high-throughput screening platform combining BV and DFT calculations. The majority of 2114 compounds crystallize in the $I\bar{4}2m$ space group as shown in **Figs.1a-c**, and their specific atomic positions can be found in **Table S1**. For these structures, distorted $LiQ_4$ tetrahedra with two different Q-Li-Q bond angles are interconnected by sharing corners (**Fig.1b**), resulting in the formation of two-dimensional (2D) layers. These layers are further bonded with segregated $MQ_4$ units, ultimately creating a three-dimensional (3D) tunnel structure within which the B cations are positioned. The $BQ_8$ dodecahedron, featuring two types of similar bond lengths of B-Q, could be visualized as the combination of two interpenetrating tetrahedra. When viewed along the a-axis, the structure reveals that the $BQ_8$ dodecahedra are stacked up by edge-sharing $MQ_4$/corner-sharing $BQ_8$ units.

In order to check the thermodynamic and kinetic stability of $Li_2BMQ_4$, we evaluated their decomposition reaction energies and phonon spectra as shown in **Table S2** and **Fig.S2**. Combining calculated decomposition energies (with positive/marginally-negative value), phonon spectra (with the absence of imaginary frequency modes) and previous experimental reports (some of 2114 compounds have been synthesized and used as promising optical materials) [35-37], we have observed that the mostly $Li_2BMQ_4$ compounds exhibit thermodynamic and kinetic stability in the $I\bar{4}2m$ space group. Based on this, we will further conduct detailed analysis of moisture stability, electronic structures and ionic transport properties for these stable 2114 compounds.

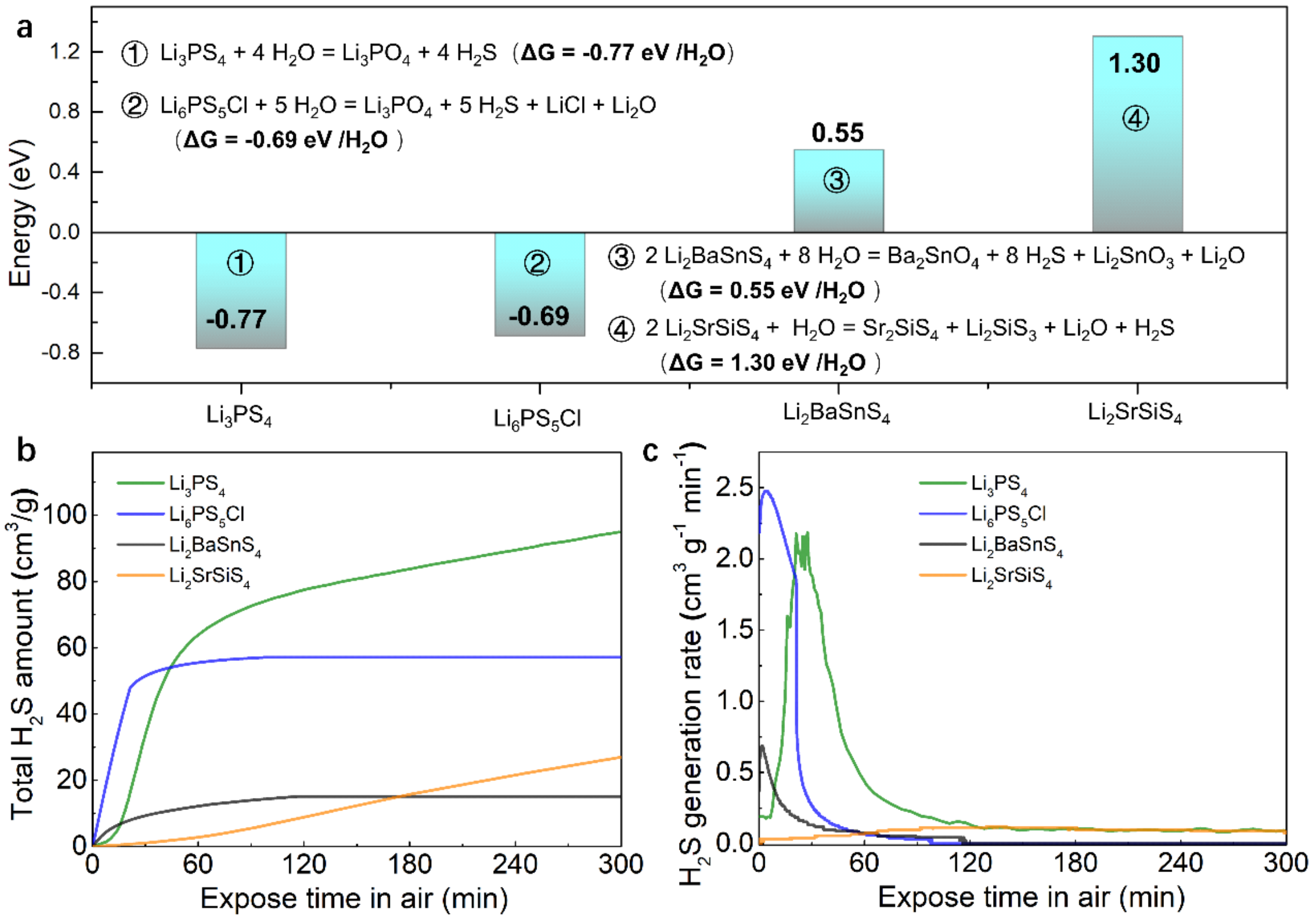

**Fig.2 The moisture stability of multiple sulfide solid electrolytes including $Li_3PS_4$, $Li_6PS_5Cl$, $Li_2BaSnS_4$ and $Li_2SrSiS_4$. (a)** The calculated hydrolysis reaction energies of four sulfides. **(b)** The total $H_2S$ generation amount of four sulfides as a function of exposure time. **(c)** The generation rate of $H_2S$ of the four sulfides, i.e., the first order derivative of curves in **Fig.2b**.

## Assessing Air Stability

As is widely recognized that the significant hurdle in the large-scale production of sulfide SEs is their poor stability against moisture in the air, hence it is crucial to evaluate the moisture stability of the novel 2114 chalcogenide systems. We firstly perform thermodynamic analyses and focus on hydrolysis reactions of generating $H_2S$/$H_2Se$ to examine the moisture stability of all sulfides and selenides. The calculation scheme for possibly representative hydrolysis reactions and reaction energy are detailed in **Table S3**. The positive reaction energy indicates that almost all sulfides and selenides are stable under humid air, which is consistent with the qualitative results reported in previous experiments (some of 2114 compounds are regarded as promising optical materials) [35-37].

Aiming to further quantitatively assess the moisture stability of the 2114 systems as SE materials, we take $Li_2BaSnS_4$ and $Li_2SrSiS_4$ as examples and compare them with common sulfide SEs, namely $Li_3PS_4$ and $Li_6PS_5Cl$. As shown in **Fig.2a**, $Li_2BaSnS_4$ and $Li_2SrSiS_4$ exhibits good moisture stability based on positive reaction energy, while $Li_3PS_4$ and $Li_6PS_5Cl$, having highly thermodynamically favorable reaction, demonstrate poor moisture stability. Furthermore, we successfully synthesized $Li_2BaSnS_4$ and $Li_2SrSiS_4$ by high temperature solid-state method [35-37], and characterize their structures by X-ray diffraction. Their crystal structures crystallized in the $I\bar{4}2m$ space group (**Fig.S3**), corresponding to previous experiments [35-37]. In addition, the total generation amount and rate of $H_2S$ are measured by gas-detection experiments [38] for these four sulfide SEs. By the end of the 300-minute exposure period, the cumulative $H_2S$ generation for $Li_3PS_4$, $Li_6PS_5Cl$, $Li_2BaSnS_4$ and $Li_2SrSiS_4$ are 94.949, 57.085, 15.080 and 26.938 $cm^3/g$ respectively, in the order of $L_3PS_4 >$ $Li_6PS_5Cl > Li_2SrSiS_4 > Li_2BaSnS_4$. The comparative analysis demonstrates that 2114 chalcogenide SEs exhibit greater moisture stability (even comparable to $Li_4SnS_4$ [38]) than $Li_3PS_4$ and $Li_6PS_5Cl$, suggesting the introduction of elemental combinations $Sr^{2+}$-$Si^{4+}$/$Ba^{2+}$-$Sn^{4+}$ inhibits the generation of $H_2S$ for 2114 SEs. This finding aligns with the empirical HSAB theory and calculations reported by Zhu and Mo [39], indicating that $Sr^{2+}$-$Si^{4+}$/$Ba^{2+}$-$Sn^{4+}$ are more resistant to moisture than $P^{5+}$. By calculating the first order derivative of total generation amount of $H_2S$ to time, the $H_2S$ generation rate can be obtained, as shown in **Fig.2c**. The $H_2S$ generation rate of $Li_3PS_4$, $Li_6PS_5Cl$ and $Li_2BaSnS_4$ increases and reaches the maximum at exposure time ≈25 min, 5 min, 2 min respectively and then continuously decreases. In contrast, the $H_2S$ generation rate of $Li_2SrSiS_4$ start to rise only within the first ≈30 min, followed by a slow increase continuing until reaching the maximum at time ≈120 min, indicating that it begins to react with water only after 30 minutes. In general, $Li_2BaSnS_4$ and $Li_2SrSiS_4$ show the most robust moisture stability among these four sulfide SEs, implying the air stability feature within the 2114 systems.

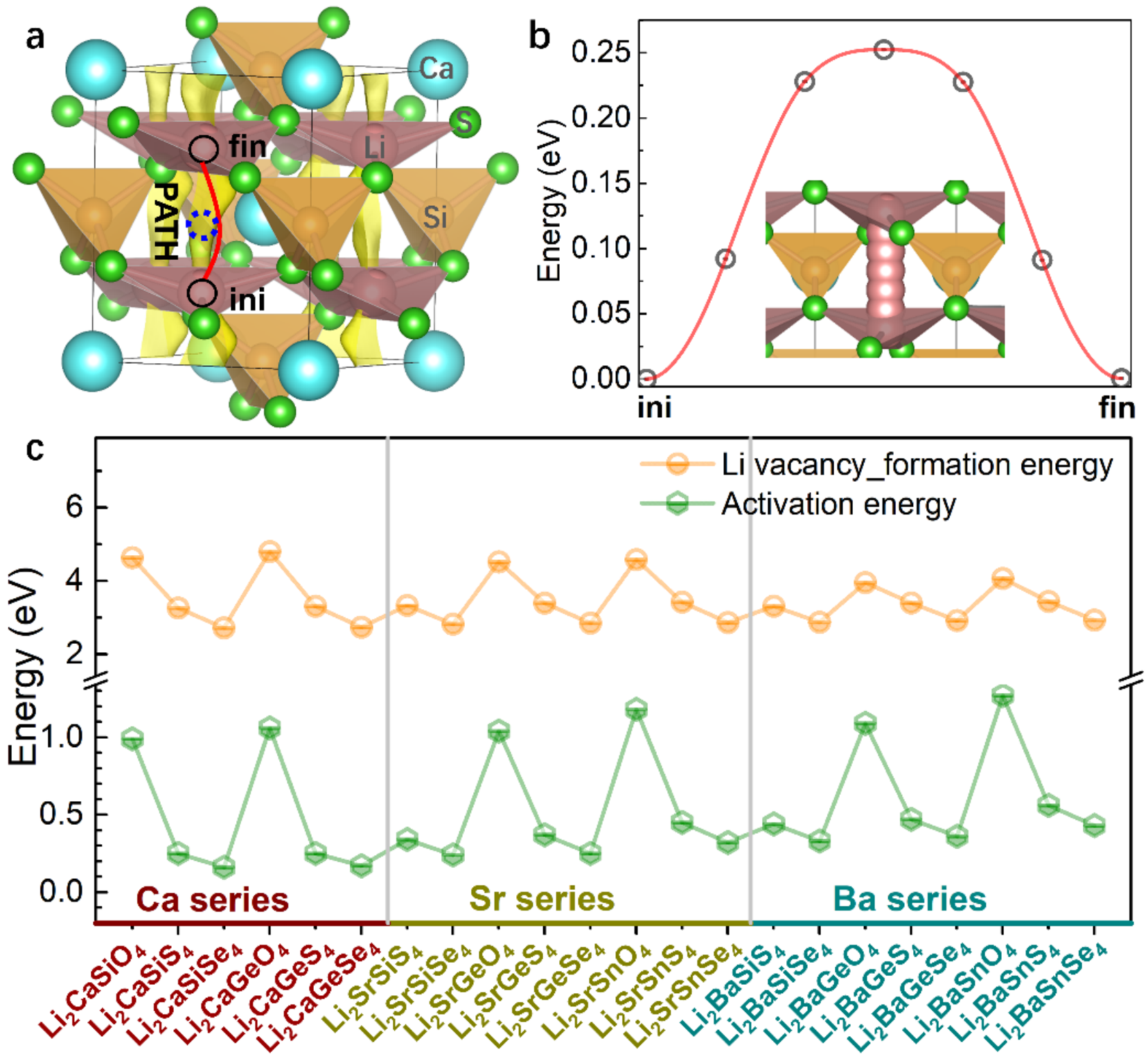

**Fig.3 The calculated $Li^+$ migration kinetic properties of $Li_2BMQ_4$ systems. (a)** The continuous $Li^+$ migration channels determined by BV method, **(b)** the energy barrier simulated by NEB method for $Li_2CaSiS_4$. **(c)** The formation energy of $Li^+$ vacancies and the activation energy of Li ions for 2114 compounds thermodynamically and kinetically stabilized in the $I\bar{4}2m$ space group.

## Ion Transport Mechanisms and Performance

Currently, we have confirmed the thermodynamic, kinetic and moisture stability of these sulfides and selenides for 2114 systems, the subsequent focus will be on investigating their lithium-ion diffusion properties. $Li_2CaSiS_4$ is selected for detailed research by combing BV method and DFT simulation in addition with the Nudged Elastic Band (NEB) method. First, we use quasi-empirical BV method to acquire the possible ionic transportation channels and interstitial sites, where the iso-surfaces of potential energies (marked by yellow belts in **Fig.3a**) are regarded as the continuous network for $Li^+$ ion transport. An interstitial 8h site (0, 0.5, $z$=0.5) related to $Li^+$ migration is signed with blue dotted circles, which is located in the continuous $Li^+$ migration pathway and connected to mobile $Li^+$ ions between two adjacent Li-S layers along *c* axis. The favorite $Li^+$ ion migration pathway suggests one-dimensional transportation behavior should be expected in these compounds and the fast ionic conducting channel is along the *c*-axis of the crystal structure.

In order to understand the kinetic properties revealed by the BV method, NEB calculations are carried out to obtain activation energy for the $Li^+$ ion hopping pathway. For $Li_2CaSiS_4$, the calculated activation energy $E_a$ with a value of 0.25 eV and corresponding barrier shape is shown in **Fig.3b**. In addition, we also calculate the activation energy for all other 2114 compounds stabilized in the $I\bar{4}2m$ space group using NEB method. As shown in **Fig.3c**, the calculated activation energies of all sulfide- and selenide-based compounds are between 0.16 eV and 0.56 eV, which are considerably comparable to the generally lithium superior conductors such as $Li_{10}GeP_2S_{12}$ (LGPS, 0.24 eV) [13], garnet $Li_7La_3Zr_2O_{12}$ (0.31 eV) [40], whereas all 2114 oxides have a higher activation energy ~ 1.0 eV. With an ultralow energy barrier (0.16 ~ 0.56 eV), these 2114 sulfides and selenides are highly expected to be prominent fast ionic conductors. To achieve high ionic conductivity, a low activation energy and a high concentration of mobile ion carriers (such as vacancies or interstitials) are essential [41]. Furthermore, we have assessed the formation energy of Li vacancies in all thermodynamically and kinetically stable 2114 compounds, revealing a range of 2.7 to 4.8 eV, which is much higher than activation energies. This implies that the elevated Li vacancy formation energy may impede the overall $Li^+$ conductivity, which will be discussed in the following section.

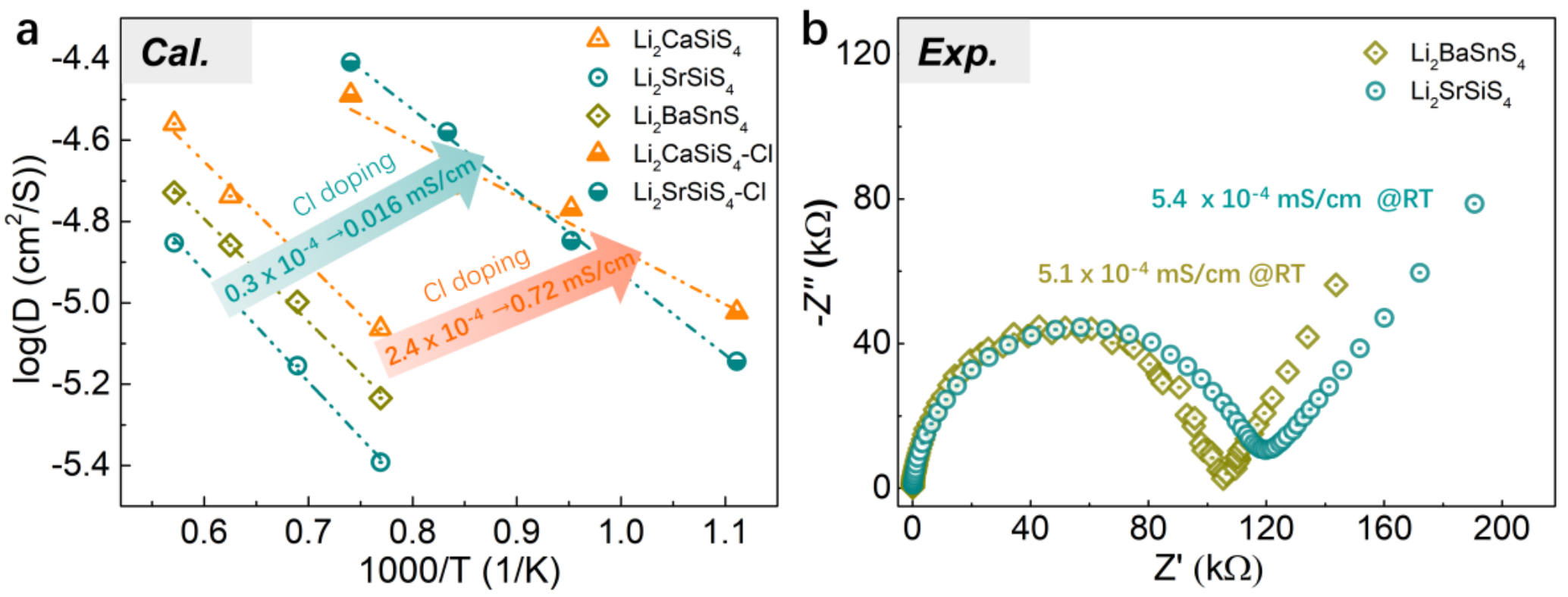

| Compounds | $Li_2CaSiS_4$ | $Li_2BaSnS_4$ | $Li_2SrSiS_4$ | $Li_2CaSiS_4$-Cl | $Li_2SrSiS_4$-Cl |
|---|---|---|---|---|---|
| ***Cal.*** $E_a$(eV) | 0.49 | 0.50 | 0.54 | 0.26 | 0.39 |
| ***Cal.*** $\sigma_T$(mS/cm) | 2.4 x $10^{-4}$ | 1.2 x $10^{-4}$ | 0.3 x $10^{-4}$ | 0.72 | 0.016 |
| ***Exp.*** $\sigma_T$(mS/cm) | -- | 5.1 x $10^{-4}$ | 5.4 x $10^{-4}$ | -- | -- |

**Fig.4 The ionic conductivity of lithium ion for 2114 compounds. (a)** The calculated Arrhenius plots of $Li^+$ diffusivity D as a function of temperature T in $Li_2CaSiS_4$, $Li_2SrSiS_4$, $Li_2BaSnS_4$, $Li_{1.75}CaSiS_{3.75}Cl_{0.25}$ ($Li_2CaSiS_4$-Cl) and $Li_{1.75}SrSiS_{3.75}Cl_{0.25}$ ($Li_2SrSiS_4$-Cl). **(b)** The experimentally measured impedance plots of the conductivity at room temperature 300 K for $Li_2BaSnS_4$ and $Li_2SrSiS_4$. The FPMD-estimated/measured activation energies $E_a$ and the ionic conductivity at room temperature $\sigma_T$ are listed in the Table. The 'Cal.' and 'Exp.' represent FPMD Calculations and Experiments.

To characterize the total $Li^+$ conductivity of 2114 compounds, First-principles molecular dynamics simulations at different temperatures $T$ (1300 K ~ 1750 K) are carried out to obtain the Arrhenius relation of $Li^+$ diffusivity as a function of $T$, which can be used to estimate the total activation energy and ionic conductivity. We take $Li_2CaSiS_4$, $Li_2BaSnS_4$ and $Li_2SrSiS_4$ as examples. The linear MSD-Δt dependence (**Fig.S5**) is utilized to apply the Einstein relation for inferring the diffusivity, while the Arrhenius equation is employed for linearly fitting log(D) against 1/T to derive the total activation energy and ionic conductivity. As shown in **Fig.4**, the fitted activation energy $E_a$ and extrapolated $Li^+$ conductivity $\sigma_T$ at 300 K is about 0.5 eV and $10^{-5}$~$10^{-4}$ mS/cm for $Li_2CaSiS_4$, $Li_2BaSnS_4$ and $Li_2SrSiS_4$ respectively. Noting that the determined activation energy by FPMD calculations for $Li_2CaSiS_4$ and $Li_2SrSiS_4$ is approximately 0.2 eV higher than that obtained from NEB calculations, whereas the activation energy of $Li_2BaSnS_4$ determined by FPMD and NEB calculations is roughly similar (**Fig.3** and **Fig.4**).

It is generally recognized that low activation energy and an abundant presence of mobile ion carriers, such as vacancies or interstitials, are crucial to achieve high ionic conductivity [41]. FPMD simulations provide a dynamic and temperature-dependent view of the activation barrier, capturing the full complexity of diffusion events, i.e., mobile ion carriers and mobile ion hops [42]. In contrast, the NEB calculations offer a more straightforward and computationally efficient approach to determine the activation barrier by directly searching for the minimum energy path on the potential energy surface [43]. For $Li_2CaSiS_4$ and $Li_2SrSiS_4$, the higher $Li^+$ vacancy formation energy (~3.3 eV in **Fig.3c**) makes it difficult to generate Li vacancy carrier, resulting in a lower concentration of vacancy carriers and reduced ionic conductivity in these two systems. Consequently, the activation energy barrier evaluated by FPMD calculations is about 0.2 eV lower than that calculated by NEB calculations. In addition, typical FPMD simulations are usually limited to a system size of a few hundred atoms and a total physical time duration of tens to hundreds of picoseconds, resulting in a limited number of diffusion events. As a result, the estimated ionic conductivity $\sigma_T$ at 300 K usually has an error bound within one order of magnitude [42]. This also explains why the FPMD-extrapolated ionic conductivity of $Li_2SrSiS_4$ (with formation energy ~3.3 eV and NEB-activation energy 0.34 eV) is about an order of magnitude lower than that of $Li_2BaSnS_4$ (with formation energy ~3.4 eV and NEB-activation energy 0.56 eV). The improved simulation approach will be discussed in the ‘Discussion’ section for our future work.

In order to validate the $Li^+$ conductivity capability of the 2114 compounds, we conduct electrochemical impedance spectroscopy measurements on $Li_2BaSnS_4$ and $Li_2SrSiS_4$ at room temperature, as shown in **Fig.4b**. The Nyquist plots show the typical features of solid fast ionic conductors, with one semi-circle and a Warburg-like tail, which is ascribed to the contribution of bulk/grain boundary resistance and the solid-state diffusion of the lithium-ion in electrode respectively. According to the total resistance obtained from the intercept on real axis or the intersection between semi-circle and Warburg-like tail, the total ionic conductivity at room temperature is estimated to be ~$5\times10^{-4}$ mS/cm at room temperature for both compounds, slightly higher or similar to the values obtained from FPMD calculations.

By comparing the results from FPMD calculations, NEB calculations and Electrochemical impedance spectroscopy measurements, we find that lithium vacancy formation energy plays a crucial role in determining the total $Li^+$ conductivity. Aiming to mitigate the inhibition of Li vacancy formation, we adopt element doping with low-valence anions to intentionally create lithium vacancies. By substituting $S^{2-}$ with $Cl^-$ in $Li_2CaSiS_4$ and $Li_2SrSiS_4$, we form $Li_{1.75}CaSiS_{3.75}Cl_{0.25}$ and $Li_{1.75}SrSiS_{3.75}Cl_{0.25}$, and perform FPMD simulations from 900 K to 1350 K (**Fig.4**). Combining Einstein and Arrhenius relations, we deduce that the activation energy of Cl-doped compounds has been reduced by approximately 0.2 eV, resulting in a three-order-of-magnitude increase of $Li^+$ conductivity at room temperature, reaching 0.016~0.72 mS/cm. This suggests that the introduction of lithium vacancies via element doping is a highly effective approach for significantly enhancing $Li^+$ conductivity. The absence of experimental measurements for $Li_2CaSiS_{3.75}Cl_{0.25}$ and $Li_2SrSiS_{3.75}Cl_{0.25}$ will be discussed below.

## Discussion

To the best of our knowledge, commonly used solid electrolytes in lithium-ion batteries, such as $Li_{10}GeP_2S_{12}$, $Li_7La_3Zr_2O_{12}$, $Li_{1.3}Al_{0.3}Ti_{1.7}(PO_4)_3$, were typically derived from element doping on their respective parent materials, including $Li_4GeS_4$-$Li_3PS_4$, $Li_5La_3M_2O_{12}$ (M = Nb, Ta) and $LiTi_2(PO_4)_3$ [45]. These doped SEs usually exhibit an enhanced $Li^+$ conductivity that is about 1-3 orders of magnitude higher than their parent materials. This is primarily attributed to the broadening ion migration channels and/or increasing the concentration of mobile ions (from element doping). Moreover, improved synthesis techniques such as solid-state reactions, sol-gel methods, etc. could further augment the ionic conductivity by mitigating grain-boundary resistance. The $Li_2BMQ_4$ series exhibit the typical features of solid fast ionic conductors, with considerable $Li^+$ ionic conductivity ~$10^{-4}$ mS/cm at room temperature. Therefore, further tailoring of the compositions by appropriate chemical substitutions and modifying synthesizing methods are crucial for the potential application of this new 2114 family of solid electrolytes in practical lithium-ion batteries. In addition to the substitution of $Cl^-$ at $S^{2-}$ site, $Al^{3+}$, $Ga^{3+}$ and $As^{5+}$ ions can also be used to replace B and M sites respectively to enhance lithium-ion conductivity as shown in **Fig.S6** of Supplementary Materials.

In this work, we aim to present the novel crystal structure of 2114 chalcogenide SEs that exhibit air stability and potentially can be used as fast ionic conductors, which have been confirmed by Electrochemical impedance spectroscopy measurements. The FPMD simulations on configurations $Li_{1.75}CaSiS_{3.75}Cl_{0.25}$ and $Li_{1.75}SrSiS_{3.75}Cl_{0.25}$ are intended to illustrate the similarity of 2114 systems to other common fast ionic conductors, which have potential for further improvement in ionic conductivity through approaches such as element doping. For configurations $Li_{1.75}CaSiS_{3.75}Cl_{0.25}$ and $Li_{1.75}SrSiS_{3.75}Cl_{0.25}$, the selection of doping elements and concentrations was currently determined through a small range of element screening and concentration testing. Beyond employing Cl element for anion doping, the possibility of cation doping and even simultaneous co-doping of both anions and cations could also be explored in the future. Managing multiple doping scenarios requires not only careful selection of element types and their concentration distributions but also consideration of element occupancy, including lattice and interstitial sites (such as 8h site), to modulate parameters such as the size of ion transport channels and the concentration of charge carriers. To achieve above goals, the simulation approach that integrates first-principles calculations with classical molecular dynamics, leveraging accelerated machine learning techniques [44], needs to be used next. It is important to note that while enhancing ionic conductivity through element doping, attention must also be paid to the air stability of the materials to achieve solid electrolytes with excellent overall performance. All in all, more efforts in experiments and calculations for the exploration of higher ionic conductivity in $Li_2BMQ_4$ systems are demanded.

## Conclusion

In summary, the $Li_2BMQ_4$ series (B = Ca, Sr and Ba; M = Si, Ge and Sn; Q = O, S and Se) represent a novel class of solid-state lithium-ion conductors characterized by a unique crystal structure ($I\bar{4}2m$ space group). With low diffusion activation energy, robust moisture stability, considerable bandgaps and ionic conductivity, this new family closely rivals promising fast ionic conductors. Leveraging past successes in regulating chemical compositions and improving synthesis methods for fast ionic conductors, similar adjustment approaches are anticipated to be applied to the $Li_2BMQ_4$ systems to achieve even higher ionic conductivity. The new family of sulfide- and selenide-based solid electrolyte shows significant potential in simultaneously possessing exceptional moisture stability and excellent ionic conductivity.

## Experimental Section

### *Calculational Methods*

In this work, we constructed crystal structures (including lattice constants, atomic coordinate and Wyckoff positions) of $Li_2BMQ_4$ (2114, B = Ca, Sr and Ba; M = Si, Ge and Sn; and Q = O, S and Se) compounds based on references [35-37, 46]. Some of $Li_2BMQ_4$ structures were obtained from the substitution of Li at Ag site in $Ag_2BMQ_4$ systems. All the calculations were carried out using Vienna Ab-initio Simulation Package (VASP) code based on density functional theory [47, 48]. In the standard DFT calculations, the exchange-correlation potential is described via the generalized gradient approximation (GGA) in the Perdew-Burke-Ernzerhof (PBE) form [49]. The calculations employed 4×4×3 k-mesh grid and energy cutoff of 520 eV. Ions were relaxed in the optimization with the energy and force convergence criterion of $10^{-5}$ eV and 0.01 eV/Å.

### *Calculational Methods--Electronic structures*

In the electronic band structure calculations, Perdew-Burke-Ernzerhof (PBE) is well-known to underestimate the bandgaps of semiconductors and insulators, whereas the modified Becke-Johnson (MBJ) exchange potential [50, 51] are known to evaluate bandgaps with an accuracy similar to hybrid functional or many-body perturbation (GW) methods, but computationally less expensive. Therefore, the MBJ exchange potential was adopted for density of states (DOS) calculations, in order to yield more accurate bandgaps.

### *Calculational Methods—thermodynamic/kinetic/air stability*

To identify the thermodynamic stability of most 2114 sulfides and selenides, we calculated the total energy of 2114 compounds and corresponding decomposition products in Table S2. The positive/negative decomposition reaction energy indicates it is possible to be thermodynamically stable/unstable and can/cannot be synthesized for this compound. To confirm the kinetic stability of 2114 compounds that were not reported experimentally, the phonon dispersion curves were calculated using the phonopy package [52] based on 2×2×2 supercell containing 128 atoms and 2×2×2 k-mesh. The kinetic stability was evaluated by the existence or absence of imaginary frequency modes in the phonon dispersion curves. In addition, we also calculated the hydrolysis reaction energies of most 2114 compounds to assess their air stability.

### *Calculational Methods—Li-ion diffusion*

We used the quasiempirical bond-valence (BV) method [53, 54] to obtain the possible ionic transportation channels and interstitial sites of these new structures. Based on the favorite Li-ion migration channel, the migration barrier of Li ion was acquired by the nudged elastic band (NEB) method as implemented in VASP [43] based on 2×2×2 supercell. Furthermore, the first-principles molecular dynamics (FPMD) was also carried out on the 2×2×2 supercell with the temperature of 900 –1750 K, lasting for 55000 steps with a time step of 1 fs. The first 6 ps are used to equilibrate the system at the temperature, and the mean-squared displacement (MSD) was calculated from the last 50 ps. Based on FPMD tests with different doping elements and doping concentrations in **Fig.S6**, we selected Cl-doped $Li_{1.75}CaSiS_{3.75}Cl_{0.25}$ and $Li_{1.75}SrSiS_{3.75}Cl_{0.25}$ for FPMD calculations at different temperatures. For the Cl-doping configurations in $Li_2CaSiS_4$ and $Li_2SrSiS_4$, we employed the enumlib program [55] to eliminate duplicate configurations, thus identifying the irreducible configurations. Through energy calculations conducted on these configurations, we ultimately pinpointed the most stable configuration with the lowest energy.

### *Experimental Methods – Synthesis and Characterization*

The $Li_2BaSnS_4$ and $Li_2SrSiS_4$ compounds were synthesized by solid-state reaction method described in reference [35-37, 46]. Powder X-ray diffraction (XRD) patterns were collected on a Bruker D2 X-ray diffractometer with Cu K$\alpha$ radiation ($\lambda$ = 1.5418 Å) at room temperature by covering the scattering angle 2θ range 10-70 degrees. The XRD patterns identify the crystal structure of these two compounds with $I\bar{4}2m$ space group, in consistent with what was reported in reference [35-37].

***Experimental Methods –Ionic conductivity and Air stability measurements***

The $Li_2BaSnS_4$ and $Li_2SrSiS_4$ powder was pressed into a pellet (diameter 10 mm; thickness 0.5 mm). The ionic conductivity of the prepared sulfides was measured by the electrochemical impedance spectroscopy (EIS) method. EIS measurements were conducted in the frequency range of 1 Hz to 8 MHz and the amplitude of 50 mV by a Zennium Pro Electrochemical Workstation. In addition, the moisture stability was measured by the specially designed equipment described in reference [39].

## Acknowledgments

This work was supported by the funding from the National Natural Science Foundation of China (grants no. 22303114, 52172258). The work was carried out at National Supercomputer Center in Tianjin, and the calculations were performed on Tianhe new generation supercomputer.

## Conflict of Interest

The authors declare no conflict of interest.

## Reference

[1] Hu, Y. S. Batteries: Getting Solid. *Nat. Energy* **2016**, *1*, 16042.

[2] Manthiram, A.; Yu, X. W.; Wang, S. F. Lithium Battery Chemistries Enabled by Solid-state Electrolytes. *Nat. Rev. Mater.* **2017**, *2*, 16103.

[3] Zhao, Q.; Stalin, S.; Zhao, C.-Z.; Archer, L. A. Designing Solid-State Electrolytes for Safe, Energy-dense Batteries. *Nat. Rev. Mater.* **2020**, *5*, 229–252.

[4] Chen, R. S.; Li, Q. H.; Yu, X. Q.; Chen, L.; Li, H. Approaching Practically Accessible Solid-State Batteries: Stability Issues Related to Solid Electrolytes and Interfaces. *Chem. Rev.* **2020**, *120*, 6820–6877.

[5] Sun, C.; Liu, J.; Gong, Y.; Wilkinson, D. P.; Zhang, J. Recent Advances in All-Solid-State Rechargeable Lithium Batteries. *Nano Energy* **2017**, *33*, 363–386.

[6] Wu, F.; Liu, L.; Wang, S.; Xu, J.; Lu, P.; Yan, W.; Peng, J.; Wu, D.; Li, H. Solid State Ionics-Selected Topics and New Directions. *Prog. Mater. Sci.* **2022**, *126*, 100921.

[7] Bachman, J. C.; Muy, S.; Grimaud, A.; Chang, H.-H.; Pour, N.; Lux, S. F.; Paschos, O.; Maglia, F.; Lupart, S.; Lamp, P.; Giordano, L.; Shao-Horn, Y. Inorganic Solid-State Electrolytes for Lithium Batteries: Mechanisms and Properties Governing Ion Conduction. *Chem. Rev.* **2016**, *116*, 140–162.

[8] Zhao, N.; Khokhar, W.; Bi, Z.; Shi, C.; Guo, X.; Fan, L.-Z.; Nan, C.-W. Solid Garnet Batteries. *Joule* **2019**, *3*, 1190–1199.

[9] Abouali, S.; Yim, C.-H.; Merati, A.; Abu-Lebdeh, Y.; Thangadurai, V. Garnet-Based Solid State Li Batteries: From Materials Design to Battery Architecture. *ACS Energy Lett.* **2021**, *6*, 1920–1941.

[10] Aono, H.; Sugimoto, E.; Sadaoka, Y.; Imanaka, N.; Adachi, G.-y. Ionic Conductivity of Solid Electrolytes Based on Lithium Titanium Phosphate. *J. Electrochem. Soc*. **1990**, *137*, 1023−1027.

[11] Bernuy-Lopez, C.; Manalastas, W.; Lopez del Amo, J. M.; Aguadero, A.; Aguesse, F.; Kilner, J. A. Atmosphere Controlled Processing of Ga-Substituted Garnets for High Li-Ion Conductivity Ceramics. *Chem. Mater.* **2014**, *26*, 3610−3617

[12] Inaguma, Y.; Chen, L.; Itoh, M.; Nakamura, T.; Uchida, T.; Ikuta, H.; Wakihara,

M. High Ionic Conductivity in Lithium Lanthanum Titanate. *Solid State Commun.* **1993**, *86*, 689−693.

[13] Kamaya, N.; Homma, K.; Yamakawa, Y.; Hirayama, M.; Kanno, R.; Yonemura, M.; Kamiyama, T.; Kato, Y.; Hama, S.; Kawamoto, K.; Mitsui, A. A Lithium Superionic Conductor. *Nat. Mater.* **2011**, *10*, 682− 686.

[14] Yamane, H.; Shibata, M.; Shimane, Y.; Junke, T.; Seino, Y.; Adams, S.; Minami, K.; Hayashi, A.; Tatsumisago, M. Crystal Structure of a Superionic Conductor, $Li_7P_3S_{11}$. *Solid State Ion.* **2007**, *178*, 1163-1167

[15] Liu, Z.; Fu, W.; Payzant, E. A.; Yu, X.; Wu, Z.; Dudney, N. J.; Kiggans, J.; Hong, K.; Rondinone, A. J.; Liang, C. Anomalous High Ionic Conductivity of Nanoporous β-$Li_3PS_4$. *J. Am. Chem. Soc.* **2013**, *135*, 975-978.

[16] Yu, C.; Eijck, L. v.; Ganapathy, S.; Wagemaker, M.; Synthesis, Structure and Electrochemical performance of the Argyrodite $Li_6PS_5Cl$ Solid Electrolyte for Li-ion Solid State Batteries. *Electrochim. Acta* **2016**, *215*, 93-99.

[17] Kanazawa, K.; Yubuchi, S.; Hotehama, C.; Otoyama, M.; Shimono, S.; Ishibashi, H.; Kubota, Y.; Sakuda, A.; Hayashi, A.; Tatsumisago, M. Mechanochemical Synthesis and Characterization of Metastable Hexagonal $Li_4SnS_4$ Solid Electrolyte. *Inorg. Chem.* **2018**, *57*, 9925–9930.

[18] Li, X.; Liang, J.; Chen, N.; Luo, J.; Adair, K. R.; Wang, C.; Banis, M. N.; Shan, T.-K.; Zhang, L.; Zhao, S.; Lu, S.; Huang, H.; Li, R.; Sun, X. Water-Mediated Synthesis of a Superionic Halide Solid Electrolyte, *Angew. Chem. Int. Ed.* **2019**, *58*, 16427–16432.

[19] Manthiram, A.; Yu, X.; Wang, S. Lithium Battery Ehemistries Enabled by Solid-State Electrolytes. *Nat. Rev. Mater*. **2017**, *2*, 16103.

[20] Zhao, S.; Zhu, X.; Jiang, W.; Ji, Z.; Ling, M.; Wang, L.; Liang, C. Fundamental Air Stability in Solid-State Electrolytes: Principles and Solutions. *Mater. Chem. Front.* **2012**, *5*, 7452–7466.

[21] Lu, P.; Wu, D.; Chen, L.; Li, H.; Wu, F. Air Stability of Solid-State Sulfide Batteries and Electrolytes, *Electrochem. Energy Rev.* **2022**, *5*, 3.

[22] Chen, X.; Guan, Z.; Chu, F.; Xue, Z.; Wu, F.; Yu, Y. Air-Stable Inorganic Solid-State Electrolytes for High Energy Density Lithium Batteries: Challenges, Strategies,

and Prospects. *InfoMat* **2022**, *4*, e12248.

[23] Yang, M.; Chen, L.; Li, H.; Wu, F. Air/Water Stability Problems and Solutions for Lithium Batteries, *Energy Mater. Adv*. **2022**, *2022*.

[24] Galven, C.; Dittmer, J.; Suard, E.; Berre, F. L.; Grosnier-Lopez, M.-P. Instability of Lithium Garnets Against Moisture. Structural Characterization and Dynamics of $Li_{7-x}H_xLa_3Sn_2O_{12}$ and $Li_{5-x}H_xLa_3Nb_2O_{12}$. *Chem. Mater*. **2012**, *24*, 3335–3345.

[25] Yow, Z. F.; Oh, Y. L.; Gu, W.; Prasada. R.; Adams, S. Effect of $Li^+/H^+$ Exchange in Water Treated Ta-doped $Li_7La_3Zr_2O_{12}$. *Solid State Ion.* **2016**, *292*, 122–129.

[26] Wang, S.; Xu, X.; Cui, C.; Zeng, C.; Liang, J.; Fu, J.; Zhang, R.; Zhai, T.; Li, H. Air Sensitivity and Degradation Evolution of Halide Solid State Electrolytes Upon Exposure. *Adv. Funct. Mater.* **2022**, *32*, 2108805.

[27] Hayashi, A.; Muramatsu, H.; Ohtomo, T.; Hama, S.; Tatsumisago, M. Improvement of Chemical Stability of $Li_3PS_4$ Glass Electrolytes by Adding $M_xO_y$ (M=Fe, Zn, and Bi) Nanoparticles. *J. Mater. Chem. A* **2013**. *1*, 6320–6326.

[28] Kimura, T.; Kato, A.; Hotehama, C.; Sakuda, A.; Hayashi, A.; Tatsumisago, M. Preparation and Characterization of Lithium Ion Conductive $Li_3SbS_4$ Glass and Glass-Ceramic Electrolytes, *Solid State Ion.* **2019**, *333*, 45–49.

[29] Liang, J.; Chen, N.; Li, X.; Li, X.; Adair, K. R.; Li, J.; Wang, C.; Yu, C.; Banis, M. N.; Zhang, L.; Zhao, S.; Lu, S.; Huang, H.; Li, R.; Huang, Y.; Sun, X. $Li_{10}Ge(P_{1-x}Sb_x)_2S_{12}$ Lithium-Ion Conductors with Enhanced Atmospheric Stability. *Chem. Mater.* **2020**, *32*, 2664–2672.

[30] Kato, Y.; Hori, S.; Saito, T.; Suzuki, K.; Hirayama, M.; Mitsui, A.; Yonemura, M.; Iba, H.; Kanno, R. High-Power All-Solid-State Batteries Using Sulfide Superionic Conductors. *Nat. Energy* **2016**, *1*, 16030.

[31] Adeli, P.; Bazak, J. D.; Park, K. H.; Kochetkov, I.; Huq, A.; Goward, G. R.; Nazar, L. F. Boosting Solid-State Diffusivity and Conductivity in Lithium Superionic Argyrodites by Halide Substitution. *Angew. Chem. Int. Ed*. **2019**, *58*, 8681−8686.

[32] Xu, H.; Cao, G.; Shen, Y.; Yu, Y.; Hu, J.; Wang, Z.; Shao, G. Enabling Argyrodite Sulfides as Superb Solid-State Electrolyte with Remarkable Interfacial Stability Against Electrodes. *Energy Environ. Mater*. **2022**, *5*, 852-864.

[33] Kaib, T.; Haddadpour, S.; Kapitein, M.; Bron, P.; Schröder, C.; Eckert, H.; Roling, B.; Dehnen, S. New Lithium Chalcogenidotetrelates, LiChT: Synthesis and Characterization of the $Li^+$-Conducting Tetralithium Ortho-Sulfdostannate $Li_4SnS_4$. *Chem. Mater.* **2012**, *24*, 2211–2219.
[34] Kuhn, A.; Holzmann, T.; Nuss, J.; Lotsch, B. V. A Facile Wet Chemistry Approach Towards Unilamellar Tin Sulfide Nanosheets from $Li_{4x}Sn_{1-x}S_2$ Solid Solutions. *J. Mater. Chem. A* **2014**, *2*, 6100–6106.
[35] Wu, K.; Zhang, B.; Yang, Z.; Pan, S. New Compressed Chalcopyrite-Like $Li_2BaM^{IV}Q_4$ ($M^{IV}$ = Ge, Sn; Q = S, Se): Promising Infrared Nonlinear Optical Materials. *J. Am. Chem. Soc.* **2017**, *139,* 14885-14888.
[36] Nian, L.; Wu, K.; He, G.; Yang, Z.; Pan, S. Effect of Element Substitution on Structural Transformation and Optical Performances in $I_2BaM^{IV}Q_4$ ($I$ = Li, Na, Cu, and Ag; $M^{IV}$ = Si, Ge, and Sn; Q = S and Se). *Inorg. Chem.* **2018**, *57*, 3434-3442.
[37] Yang, Y.; Wu, K.; Wu, X.; Zhang, B.; Gao, L. A New Family of Quaternary Thiosilicates $SrA_2SiS_4$ (A = Li, Na, Cu) as Promising Infrared Nonlinear Optical Crystals. *J. Mater. Chem. C* **2020**, *8*, 1762.
[38] Lu, P.; Liu, L.; Wang, S.; Xu, J.; Peng, J.; Yan, W.; Wang, Q.; Li, H.; Chen, L.; Wu, F. Superior All-Solid-State Batteries Enabled by a Gas-Phase-Synthesized Sulfide Electrolyte with Ultrahigh Moisture Stability and Ionic Conductivity. *Adv. Mater.* **2021**, *33*, 2100921.
[39] Zhu, Y.; Mo, Y. Materials Design Principles for Air-Stable Lithium/Sodium Solid Electrolytes. *Angew. Chem. Int. Ed.* **2020**, *59*, 17472-17476.
[40] Murugan, R.; Thangadurai, V.; Weppner, W. Fast Lithium Ion Conduction in Garnet-Type $Li_7La_3Zr_2O_{12}$. *Angew. Chem. Int. Ed.* **2007**, *46*, 7778-7781.
[41] He, X.; Zhu, Y.; Mo, Y. Origin of Fast Ion Diffusion in Super-Ionic Conductors. *Nat. Commun.* **2017**, *8*, 15893.
[42] He, X.; Zhu, Y.; Epstein, A.; Mo, Y. Statistical variances of diffusional properties from ab initio molecular dynamics simulations. *Npj Comput. Mater.* **2018**, *4(1)*, 18.
[43] Henkelman, G.; Jónsson, H. Improved Tangent Estimate in the Nudged Elastic Band Method for Finding Minimum Energy Paths and Saddle Points. *J. Chem. Phys.*

**2000**, *113*, 9978.

[44] Wang, Y.; Wu, S.; Shao, W.; Sun, X.; Wang, Q.; Xiao, R.; Li, H. Accelerated strategy for fast ion conductor materials screening and optimal doping scheme exploration. *J. Materiomics* **2022**, *8*, 1038-1047.

[45] Zheng, F.; Kotobuki, M.; Song, S.; Lai, M. O.; Lu, L. Review on Solid Electrolytes for All-Solid-State Lithium-Ion Batteries. *J. Power Sources* **2018**, *389*, 198-213.

[46] Wu, K.; Chu, Y.; Yang, Z.; Pan, S. $A_2SrM^{IV}S_4$ (A = Li, Na; $M^{IV}$ = Ge, Sn) Concurrently Exhibiting Wide Bandgaps and Good Nonlinear Optical Responses as New Potential Infrared Nonlinear Optical Materials. *Chem. Sci.* **2019**, *10*, 3963.

[47] Kresse, G.; Furthmüller, J. Efficient Iterative Schemes for Ab Initio Total-Energy Calculations Using a Plane-Wave Basis Set. *Phys. Rev. B* **1996**, *54*, 11169.

[48] Kresse, G; Furthmüller, J. Efficiency of Ab-Initio Total Energy Calculations for Metals and Semiconductors Using a Plane-Wave Basis Set. *Comput. Mater. Sci.* **1996**, *6*, 15.

[49] Perdew, J. P.; Burke, K; Ernzerhof, M. Generalized Gradient Approximation Made Simple. *Phys. Rev. Lett.* **1996**, *77*, 3865.

[50] Becke, A. D.; Johnson, E. R. A Simple Effective Potential for Exchange. *J. Chem. Phys.* **2006**, *124*, 221101.

[51] Tran, F.; Blaha, P. Accurate Band Gaps of Semiconductors and Insulators with a Semilocal Exchange-Correlation Potential. *Phys. Rev. Lett.* **2009**, *102*, 226401.

[52] Togo, A.; Tanaka, I. First Principles Phonon Calculations in Materials Science. *Scr. Mater.* **2015**, *108*, 1.

[53] Xiao, R.; Li, H.; Chen, L. High-Throughput Design and Optimization of Fast Lithium Ion Conductors by the Combination of Bond-Valence Method and Density Functional Theory. *Sci. Rep.* **2015**, *5*, 14227.

[54] Adams, S.; Rao, R. P. High Power Lithium Ion Battery Materials by Computational Design. Phys. *Status Solidi A* **2011**, *208*, 1746.

[55] Lian, J.-C.; Wu, H.-Y.; Huang, W.-Q.; Hu, W.; Huang, G.-F. Algorithm for Generating Irreducible Site-Occupancy Configurations. *Phys. Rev. B* **2020**, *102*, 134209.

# Supplementary Materials for "A family of air-stable chalcogenide solid electrolytes in $Li_2BMQ_4$ (B = Ca, Sr and Ba; M = Si, Ge and Sn; Q = O, S and Se) systems"

Huican Mao[1, 2#], Xiang Zhu[2#], Guangmao Li[3#], Jie Pang[2], Junfeng Hao[2], Liqi Wang[2], Hailong Yu[2], Youguo Shi[2], Fan Wu[2*], Shilie Pan[3*], Ruijuan Xiao[2*], Hong Li[2*], Liquan Chen[2]

1. Department of Energy Storage Science and Engineering, School of Metallurgical and Ecological Engineering, University of Science and Technology Beijing, Beijing 100083, China
2. Beijing National Laboratory for Condensed Matter Physics, Institute of Physics, Chinese Academy of Sciences, 100190, Beijing, China.
3. Research Center for Crystal Materials, Xinjiang Technical Institute of Physics and Chemistry, Chinese Academy of Sciences, Urumqi 830011, China.

To explore the thermodynamic stability of $Li_2BMQ_4$ (2114, B = Ca, Sr and Ba; M = Si, Ge and Sn; Q = O, S and Se) systems, decomposition reaction energies are carried out for all sulfides and selenides. The decomposition reaction equations and decomposition reaction energies are demonstrated in **Table S2**, in which the positive/small negative value of decomposition energies indicate that these compounds can exist as stable/metastable (or be stabilized by entropic effects) structures, verifying the thermodynamic stability of these new structures. We also assessed the kinetic stability by phonon spectrum calculations for all 2114 compounds except for those experimentally reported compounds marked by small white circles inside larger brown filled shapes (diamond, triangle and pentagon) and small white circles inside larger yellow filled shapes in **Fig.S1**, which represent the compounds that have been synthesized experimentally with $I\bar{4}2m$ space group and other space groups ($P6_3cm$, $P3_121$, etc) respectively. The absence of imaginary frequency modes in the Brillouin zone (**Fig.S2**) confirms the kinetic stability of these compounds, denoted by brown filled shapes. In addition, those kinetically unstable structures, with imaginary frequency modes in phonon dispersion, are marked by white unfilled shapes. In the $Li_2BMQ_4$ systems, only five of them ($Li_2SrSiO_4$, $Li_2BaSiO_4$, $Li_2CaSnO_4$, $Li_2CaSnS_4$ and $Li_2CaSnSe_4$) are not crystallized/stable in $I\bar{4}2m$ space group, which may be attributed to the special combinations between smallest $Ca^{2+}$ and largest $Sn^{4+}$ ionic radii, smallest $Si^{4+}$/$O^{2-}$and larger $Sr^{2+}$ and $Ba^{2+}$ ionic radii within the same main group, further leading to the deviation from the structural tolerance factors with $I\bar{4}2m$ symmetry [1].

**Table S1** Take $Li_2SrSiS_4$ as an example to show the lattice constants, atomic coordinates, equivalent isotropic displacement parameters for $Li_2BMQ_4$ (B = Ca, Sr and Ba; M = Si, Ge and Sn; Q = O, S and Se) systems with $I\bar{4}2m$ space group.

| $Li_2SrSiS_4$ | Space group | | a | c |
|---|---|---|---|---|
| | $I\bar{4}2m$ | | 6.5133 | 7.8170 |
| Atoms | Wyckoff positions | x | y | z |
| Li | 4*d* | 0.00 | 0.50 | 0.25 |
| Sr | 2*a* | 0.00 | 0.00 | 0.00 |
| Si | 2*b* | 0.00 | 0.00 | 0.50 |
| S | 8*i* | 0.1905 | 0.1905 | 0.3429 |

**Table S2** The crystal structures, experimental synthesis, phonon spectra, bandgap (obtained from the density of states in **Fig.S4**) and activation energies for all $Li_2BMQ_4$ (B = Ca, Sr and Ba; M = Si, Ge and Sn; Q = O, S and Se) compounds. The positive/negative decomposition reaction energy ($E_d$) indicates it is possible to be thermodynamically stable/unstable and can/cannot be synthesized for this compound. Although the decomposition reaction energies of several Ca-based compounds are negative, the relatively small negative values imply that they can exist as a metastable structure or be stabilized by entropic effects. The absence/existence of imaginary frequency modes in the phonon dispersion curves represents that the phonon spectrum is 'Stable/Unstable'. The calculations of decomposition reaction energy (phonon spectrum) are performed for 2114 sulfides and selenides (for those compounds that are not synthesized experimentally).

| Compounds | Space group | Synthesized experimentally | Phonon spectrum | $E_d$ (meV/atom) | Bandgap (eV) | Activation energy (eV) |
|---|---|---|---|---|---|---|
| $Li_2CaSiO_4$ | $I\bar{4}2m$ | Yes[2] | -- | -- | 7.1 | 0.99 |
| $Li_2CaSiS_4$ | $I\bar{4}2m$ | No | Stable | -1.7 | 3.3 | 0.25 |
| $Li_2CaSiSe_4$ | $I\bar{4}2m$ | No | Stable | 1.7 | 2.3 | 0.16 |
| $Li_2CaGeO_4$ | $I\bar{4}2m$ | Yes[3] | -- | -- | 5.7 | 1.06 |
| $Li_2CaGeS_4$ | $I\bar{4}2m$ | No | Stable | -12.7 | 2.9 | 0.25 |
| $Li_2CaGeSe_4$ | $I\bar{4}2m$ | No | Stable | -2.5 | 1.7 | 0.17 |
| $Li_2CaSnO_4$ | Uncertain | Yes[4] | -- | -- | -- | -- |
| $Li_2CaSnS_4$ | -- | No | Unstable | -- | -- | -- |
| $Li_2CaSnSe_4$ | -- | No | Unstable | -- | -- | -- |
| $Li_2SrSiO_4$ | $P3_121$ | Yes[2] | -- | -- | -- | -- |
| $Li_2SrSiS_4$ | $I\bar{4}2m$ | Yes[5] | -- | 40.0 | 3.5 | 0.34 |
| $Li_2SrSiSe_4$ | $I\bar{4}2m$ | No | Stable | 41.9 | 2.4 | 0.24 |
| $Li_2SrGeO_4$ | $I\bar{4}2m$ | Yes[6] | -- | -- | 5.6 | 1.04 |
| $Li_2SrGeS_4$ | $I\bar{4}2m$ | Yes[7] | -- | 27.2 | 3.0 | 0.37 |
| $Li_2SrGeSe_4$ | $I\bar{4}2m$ | No | Stable | 41.4 | 1.8 | 0.25 |
| $Li_2SrSnO_4$ | $I\bar{4}2m$ | No | Stable | -- | 4.0 | 1.18 |
| $Li_2SrSnS_4$ | $I\bar{4}2m$ | Yes[7] | -- | 33.4 | 3.1 | 0.45 |
| $Li_2SrSnSe_4$ | $I\bar{4}2m$ | No | Stable | 29.5 | 2.1 | 0.32 |
| $Li_2BaSiO_4$ | $P6_3cm$ | Yes[2] | -- | -- | -- | -- |
| $Li_2BaSiS_4$ | $I\bar{4}2m$ | No | Stable | 26.5 | 3.5 | 0.44 |
| $Li_2BaSiSe_4$ | $I\bar{4}2m$ | Yes[8] | -- | 38.7 | 2.5 | 0.33 |
| $Li_2BaGeO_4$ | $I\bar{4}2m$ | No | Stable | -- | 5.2 | 1.09 |
| $Li_2BaGeS_4$ | $I\bar{4}2m$ | Yes[9] | -- | 21.7 | 3.1 | 0.47 |
| $Li_2BaGeSe_4$ | $I\bar{4}2m$ | Yes[9] | -- | 42.4 | 2.0 | 0.36 |
| $Li_2BaSnO_4$ | $I\bar{4}2m$ | No | Stable | -- | 3.8 | 1.27 |
| $Li_2BaSnS_4$ | $I\bar{4}2m$ | Yes[9] | -- | 43.9 | 3.2 | 0.56 |
| $Li_2BaSnSe_4$ | $I\bar{4}2m$ | Yes[9] | -- | 42.3 | 2.1 | 0.43 |

Decomposition reaction equations:

$Li_2CaSiS_4 = CaS + Li_2SiS_3$

$Li_2CaGeS_4 = CaS + Li_2GeS_3$

$Li_2CaGeSe_4=CaSe+GeSe_2+Li_2Se$

$2\ Li_2SrSiS_4 = Li_2S + Li_2SiS_3 + Sr_2SiS_4$

$2\ Li_2SrSiSe_4 = SiSe_2 + 2\ Li_2Se + Sr_2SiSe_4$

$2\ Li_2SrGeS_4 = Li_4GeS_4 + Sr_2GeS_4$

$2\ Li_2SrGeSe_4 = GeSe_2 + 2\ Li_2Se + Sr_2GeSe_4$

$Li_2SrSnS_4 = Li_2SnS_3 + SrS$

$Li_2SrSnSe_4 = Li_2Se + SrSe + SnSe_2$

$2\ Li_2BaSiS_4 = Ba_2SiS_4 + Li2S + Li_2SiS_3$

$2\ Li_2BaSiSe_4 = Ba_2SiSe_4 + 2\ Li_2Se + SiSe_2$

$2\ Li_2BaGeS_4 = Ba_2GeS_4 + Li_4GeS_4$

$2\ Li_2BaGeSe_4 = Ba_2GeSe_4 + 2\ Li_2Se + GeSe_2$

$2\ Li_2BaSnS_4 = Ba_2SnS_4 + Li_2S + Li_2SnS_3$

$2\ Li_2BaSnSe_4 = Ba_2SnSe_4 + 2\ Li_2Se + SnSe_2$

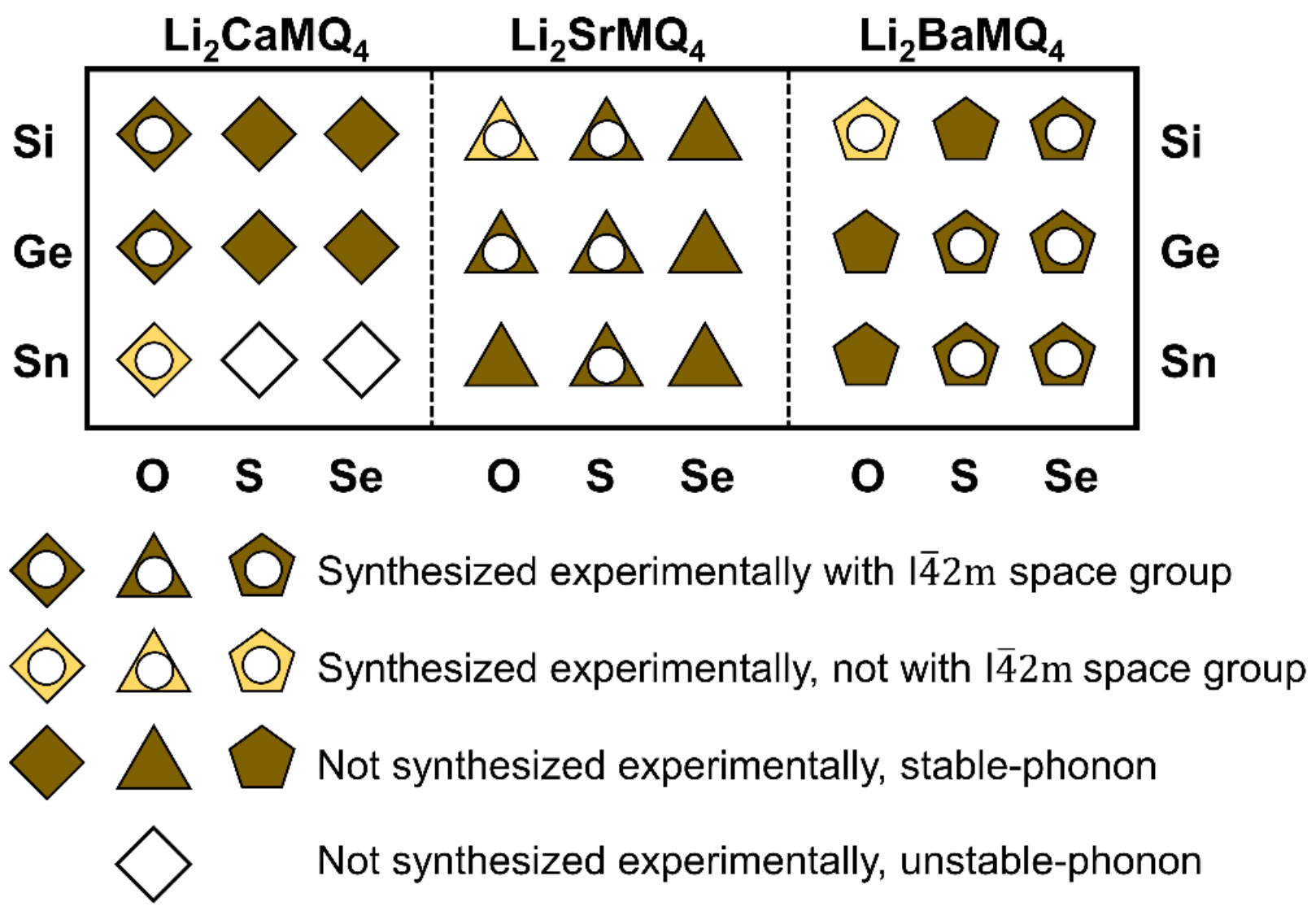

**Fig.S1** The schematic diagram of the structure stability for $Li_2BMQ_4$ (B = Ca, Sr and Ba; M = Si, Ge and Sn; Q = O, S and Se) compounds. Small white circles inside the larger brown filled shapes (diamond, triangle and pentagon) and small white circles inside the larger yellow filled shapes indicate that the compounds have been synthesized experimentally with $I\bar{4}2m$ space group and other space groups ($P6_3cm$, $P3_121$, etc.), respectively. Brown filled shapes and white unfilled shapes represent the compounds that are not synthesized experimentally, where the imaginary frequency

mode is absent in the calculated phonon dispersion curves for the materials denoted by brown filled shapes while the compounds marked by white unfilled diamond behave in an opposite manner.

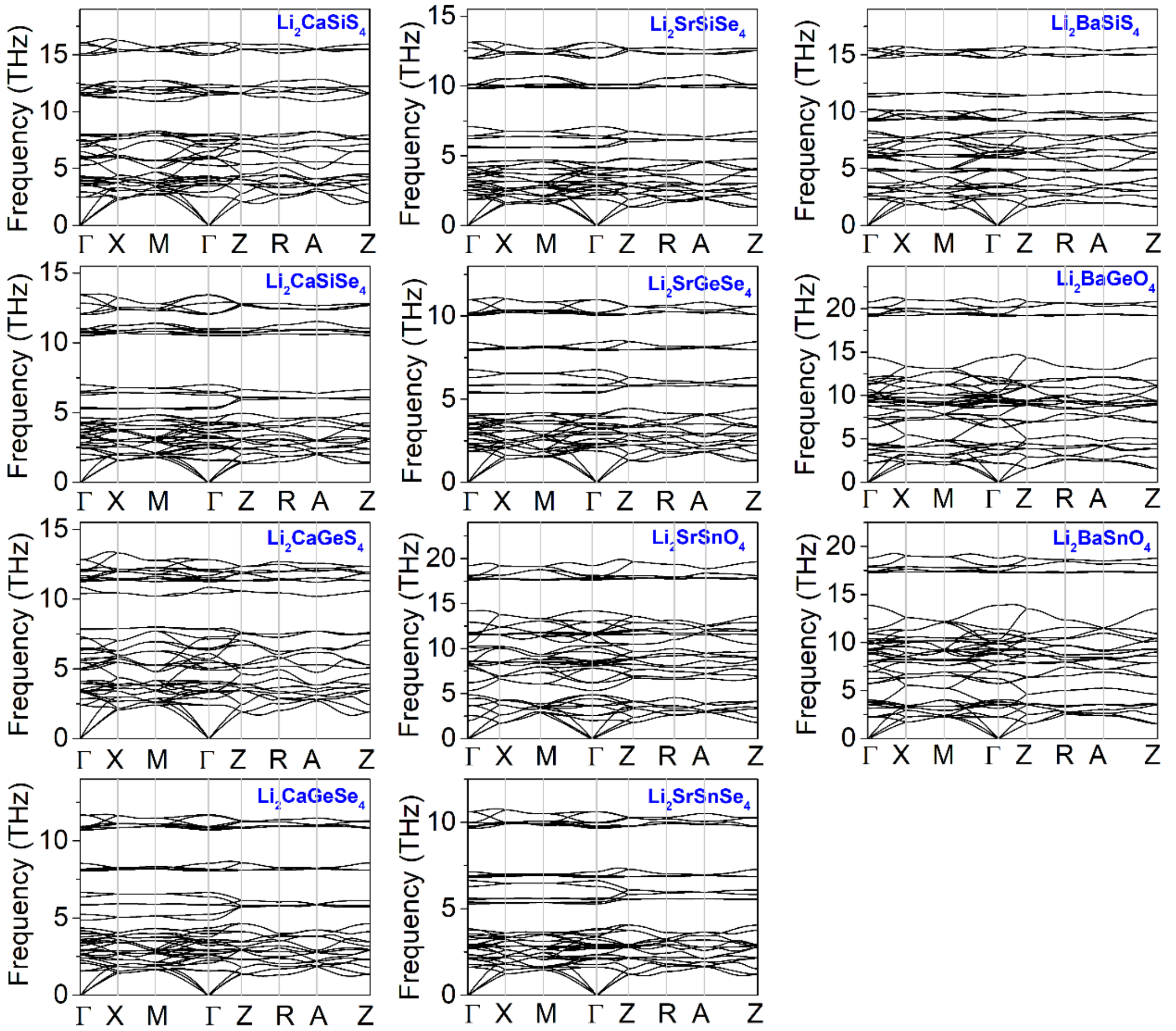

**Fig.S2** The calculated phonon dispersion curves in $Li_2BMQ_4$ (B = Ca, Sr and Ba; M = Si, Ge and Sn and Q = O, S and Se) systems except for those experimentally synthesized and phonon-unstable compounds. The absence of imaginary frequency modes suggests these compounds are kinetically stable.

**Table S3** The hydrolysis reaction energies of 2114 chalcogenides solid electrolytes (SEs) evaluated by DFT methods. More-negative reaction energy indicates worse moisture stability (moisture sensitive), whereas more-positive reaction energy indicative better moisture stability (moisture stable).

| Compounds | Hydrolysis reaction | Reaction energy ($eV/H_2O$) |
|---|---|---|
| $Li_2CaSiS_4$ | $Li_2CaSiS_4 + H_2O = CaO + Li_2SiS_3 + H_2S$ | 0.52796 |
| | $Li_2CaSiS_4 + 4\ H_2O = Li_2SiO_3 + CaO + 4\ H_2S$ | -0.45626 |
| | $Li_2CaSiS_4 + 3\ H_2O = CaS + Li_2SiO_3 + 3\ H_2S$ | -0.78895 |
| | $Li_2CaSiS_4 + 4\ H_2O = 4\ H_2S + Li_2CaSiO_4$ | -0.53477 |
| $Li_2CaSiSe_4$ | $Li_2CaSiSe_4 + H_2O = CaSe + Li_2O + H_2Se + SiSe2$ | 0.95244 |
| | $Li_2CaSiSe_4 + 5\ H_2O = CaO + SiO_2 + 4\ H_2Se + 2\ LiOH$ | -0.22620 |
| | $Li_2CaSiSe_4 + 4\ H_2O = CaO + SiO_2 + 4\ H_2Se + Li_2O$ | 0.03631 |
| | $Li_2CaSiSe_4 + 4\ H_2O = 4\ H_2Se + Li_2CaSiO_4$ | -0.37895 |
| $Li_2CaGeS_4$ | $Li_2CaGeS_4 + 4\ H_2O = 4\ H_2S + Li_2CaGeO_4$ | 0.05151 |
| | $Li_2CaGeS_4 + 4\ H_2O = Li_2GeO_3 + CaO + 4\ H_2S$ | 0.11849 |
| $Li_2CaGeSe_4$ | $Li_2CaGeSe_4 + 4\ H_2O = 4\ H_2Se + Li2CaGeO4$ | 0.32803 |
| | $Li_2CaGeSe_4 + 5\ H_2O = CaO + GeO_2 + 4\ H_2Se + 2\ LiOH$ | 0.36674 |
| | $Li_2CaGeSe_4 + 4\ H_2O = CaO + GeO_2 + 4\ H_2Se + Li_2O$ | 0.77749 |
| | $Li_2CaGeSe_4 + H_2O = CaSe + Li_2O + H_2Se + GeSe_2$ | 0.94510 |
| $Li_2CaSnS_4$ | $Li_2CaSnS_4 + 4\ H_2O = 4\ H_2S + Li_2CaSnO_4$ | 0.34419 |
| | $Li_2CaSnS_4 + 4\ H_2O = Li_2SnO_3 + CaO + 4\ H_2S$ | 0.24295 |
| $Li_2CaSnSe_4$ | $Li_2CaSnSe_4 + 4\ H_2O = 4\ H_2Se + Li_2CaSnO_4$ | 0.66934 |
| | $Li_2CaSnSe_4 + 4\ H_2O = CaO + SnO_2 + 4\ H_2Se + Li_2O$ | 0.76766 |
| $Li_2SrSiS_4$ | $2\ Li_2SrSiS_4 + H_2O = Li_2SiS_3 + Sr_2SiS_4 + Li_2O + H_2S$ | 1.30316 |
| | $2\ Li_2SrSiS_4 + 9\ H_2O = Li_2SiO_3 + Sr_2SiO_4 + 8\ H_2S + 2\ LiOH$ | -0.41345 |
| | $2\ Li_2SrSiS_4 + 8\ H_2O = Li_2SiO_3 + Sr_2SiO_4 + 8\ H_2S + Li_2O$ | -0.30560 |
| | $2\ Li_2SrSiS_4 + 4\ H_2O = Li_2SiO_3 + Sr_2SiS_4 + Li_2O + 4\ H_2S$ | -0.26246 |
| | $2\ Li_2SrSiS_4 + 5\ H_2O = Li_2SiS_3 + Sr_2SiO_4 + Li_2O + 5\ H_2S$ | -0.01835 |
| | $Li_2SrSiS_4 + 4\ H_2O = 4\ H_2S + Li_2SrSiO_4$ | -0.39931 |
| $Li_2SrSiSe_4$ | $2\ Li_2SrSiSe_4 + 2\ H_2O = Sr_2SiSe_4 + SiSe_2 + 2\ Li_2O + 2\ H_2Se$ | 1.27414 |
| | $2\ Li_2SrSiSe_4 + 10\ H_2O = Sr_2SiO_4 + SiO_2 + 8\ H_2Se + 4\ LiOH$ | -0.22775 |
| | $2\ Li_2SrSiSe_4 + 8\ H_2O = Sr_2SiO_4 + SiO_2 + 8\ H_2Se + 2\ Li_2O$ | 0.03438 |
| | $Li_2SrSiSe_4 + 4\ H_2O = 4\ H_2Se + Li_2SrSiO_4$ | -0.22771 |
| $Li_2SrGeS_4$ | $Li_2SrGeS_4 + 4\ H_2O = 4\ H_2S + Li_2SrGeO_4$ | 0.18674 |
| | $2\ Li_2SrGeS_4 + 8\ H_2O = Li_4GeO_4 + Sr_2GeO_4 + 8\ H_2S$ | 0.23063 |
| | $2\ Li_2SrGeS_4 + 4\ H_2O = Li_4GeO_4 + Sr_2GeS_4 + 4\ H_2S$ | 0.18689 |
| $Li_2SrGeSe_4$ | $Li_2SrGeSe_4 + 4\ H_2O = 4\ H_2Se + Li_2SrGeO_4$ | 0.47388 |
| | $2\ Li_2SrGeSe_4 + 10\ H_2O = Sr_2GeO_4 + GeO_2 + 8\ H_2Se + 4\ LiOH$ | 0.36929 |
| | $2\ Li_2SrGeSe_4 + 8\ H_2O = Sr_2GeO_4 + GeO_2 + 8\ H_2Se + 2\ Li_2O$ | 0.78067 |
| | $2\ Li_2SrGeSe_4 + 2\ H_2O = Sr_2GeSe_4 + GeSe_2 + 2\ Li_2O + 2\ H_2Se$ | 1.27020 |
| $Li_2SrSnS_4$ | $Li_2SrSnS_4 + 4\ H_2O = 4\ H_2S + Li_2SrSnO_4$ | 0.48324 |
| | $Li_2SrSnS_4 + 4\ H_2O = Li_2SnO_3 + SrO + 4\ H_2S$ | 0.46845 |
| | $Li_2SrSnS_4 + H_2O = Li_2SnS_3 + SrO + H_2S$ | 1.27435 |
| | $Li_2SrSnS_4 + 3\ H_2O = Li_2SnO_3 + SrS + 3\ H_2S$ | 0.28625 |
| $Li_2SrSnSe_4$ | $Li_2SrSnSe_4 + 4\ H_2O = 4\ H_2Se + Li_2SrSnO_4$ | 0.82683 |
| | $Li_2SrSnSe_4 + 5\ H_2O = SrO + SnO_2 + 4\ H_2Se + 2\ LiOH$ | 0.55403 |
| | $Li_2SrSnSe_4 + 4\ H_2O = SrO + SnO_2 + 4\ H_2Se + Li_2O$ | 1.01161 |
| | $Li_2SrSnSe_4 + H_2O = SrSe + SnSe_2 + Li_2O + H_2Se$ | 1.17463 |
| $Li_2BaSiS_4$ | $2\ Li_2BaSiS_4 + H_2O = Ba_2SiS_4 + Li_2SiS_3 + H_2S + Li_2O$ | 1.08640 |
| | $Li_2BaSiS_4 + 4\ H_2O = 4\ H_2S + Li_2BaSiO_4$ | -0.22865 |
| | $2\ Li_2BaSiS_4 + 9\ H_2O = Ba_2SiO_4 + 8\ H_2S + Li_2SiO_3 + 2\ LiOH$ | -0.35080 |
| | $2\ Li_2BaSiS_4 + 8\ H_2O = Ba_2SiO_4 + 8\ H_2S + Li_2SiO_3 + Li_2O$ | -0.23512 |
| | $2\ Li_2BaSiS_4 + 5\ H_2O = Li_2SiS_3 + Ba_2SiO_4 + Li_2O + 5\ H_2S$ | 0.09442 |
| | $2\ Li_2BaSiS_4 + 4\ H_2O = Li_2SiO_3 + Ba_2SiS_4 + Li_2O + 4\ H_2S$ | -0.31666 |
| $Li_2BaSiSe_4$ | $2\ Li_2BaSiSe_4 + 2\ H_2O = Ba_2SiSe_4 + 2\ Li_2O + 2\ H_2Se + SiSe_2$ | 1.24877 |
| | $2\ Li_2BaSiSe_4 + 10\ H_2O = Ba_2SiO_4 + SiO_2 + 8\ H_2Se + 4\ LiOH$ | -0.14684 |
| | $2\ Li_2BaSiSe_4 + 8\ H_2O = Ba_2SiO_4 + SiO_2 + 8\ H_2Se + 2\ Li_2O$ | 0.13552 |
| | $Li_2BaSiSe_4 + 4\ H_2O = 4\ H_2Se + Li_2BaSiO_4$ | -0.02639 |
| $Li_2BaGeS_4$ | $Li_2BaGeS_4 + 4\ H_2O = 4\ H_2S + Li_2BaGeO_4$ | 0.34437 |
| | $2\ Li_2BaGeS_4 + 8\ H_2O = Ba_2GeO_4 + Li_4GeO_4 + 8\ H_2S$ | 0.33248 |
| | $2\ Li_2BaGeS_4 + 4\ H_2O = Ba_2GeS_4 + Li_4GeO_4 + 4\ H_2S$ | 0.16490 |
| | $2\ Li_2BaGeS_4 + 4\ H_2O = Ba_2GeO_4 + Li_4GeS_4 + 4\ H_2S$ | 0.58690 |
| $Li_2BaGeSe_4$ | $Li_2BaGeSe_4 + 4\ H_2O = 4\ H_2Se + Li_2BaGeO_4$ | 0.66032 |
| | $2\ Li_2BaGeSe_4 + 10\ H_2O = Ba_2GeO_4 + GeO_2 + 8\ H_2Se + 4\ LiOH$ | 0.47382 |
| | $2\ Li_2BaGeSe_4 + 8\ H_2O = Ba_2GeO_4 + GeO_2 + 8\ H_2Se + 2\ Li_2O$ | 0.91134 |
| | $2\ Li_2BaGeSe_4 + 2\ H_2O = Ba_2GeSe_4 + 2\ Li_2O + 2\ H_2Se + GeSe_2$ | 1.27870 |
| $Li_2BaSnS_4$ | $Li_2BaSnS_4 + 4\ H_2O = 4\ H_2S + Li_2BaSnO_4$ | 0.63734 |
| | $2\ Li_2BaSnS_4 + 9\ H_2O = Ba_2SnO_4 + 8\ H_2S + Li_2SnO_3 +$ | 0.34673 |
| | 2 LiOH | |
| | 2 $Li_2BaSnS_4$ + 8 $H_2O$ = $Ba_2SnO_4$ + 8 $H_2S$ + $Li_2SnO_3$ + $Li_2O$ | 0.54961 |
| | 2 $Li_2BaSnS_4$ + $H_2O$ = $Ba_2SnS_4$ + $Li_2O$ + $H_2S$ + $Li_2SnS_3$ | 1.36508 |
| $Li_2BaSnSe_4$ | $Li_2BaSnSe_4$ + 4 $H_2O$ = 4 $H_2Se$ + $Li_2BaSnO_4$ | 1.00761 |
| | 2 $Li_2BaSnSe_4$ + 10 $H_2O$ = $Ba_2SnO_4$ + $SnO_2$ + 8 $H_2Se$ + 4 LiOH | 0.56047 |
| | 2 $Li_2BaSnSe_4$ + 8 $H_2O$ = $Ba_2SnO_4$ + $SnO_2$ + 8 $H_2Se$ + 2 $Li_2O$ | 1.01966 |
| | 2 $Li_2BaSnSe_4$ + 9 $H_2O$ = $Ba_2SnO_4$ + $Li_2SnO_3$ + 8 $H_2Se$ + 2 LiOH | 0.67586 |
| | 2 $Li_2BaSnSe_4$ + 8 $H_2O$ = $Ba_2SnO_4$ + $Li_2SnO_3$ + 8 $H_2Se$ + $Li_2O$ | 0.91988 |
| | 2 $Li_2BaSnSe_4$ + 2 $H_2O$ = $Ba_2SnSe_4$ + 2 $Li_2O$ + 2 $H_2Se$ + $SnSe_2$ | 1.27748 |

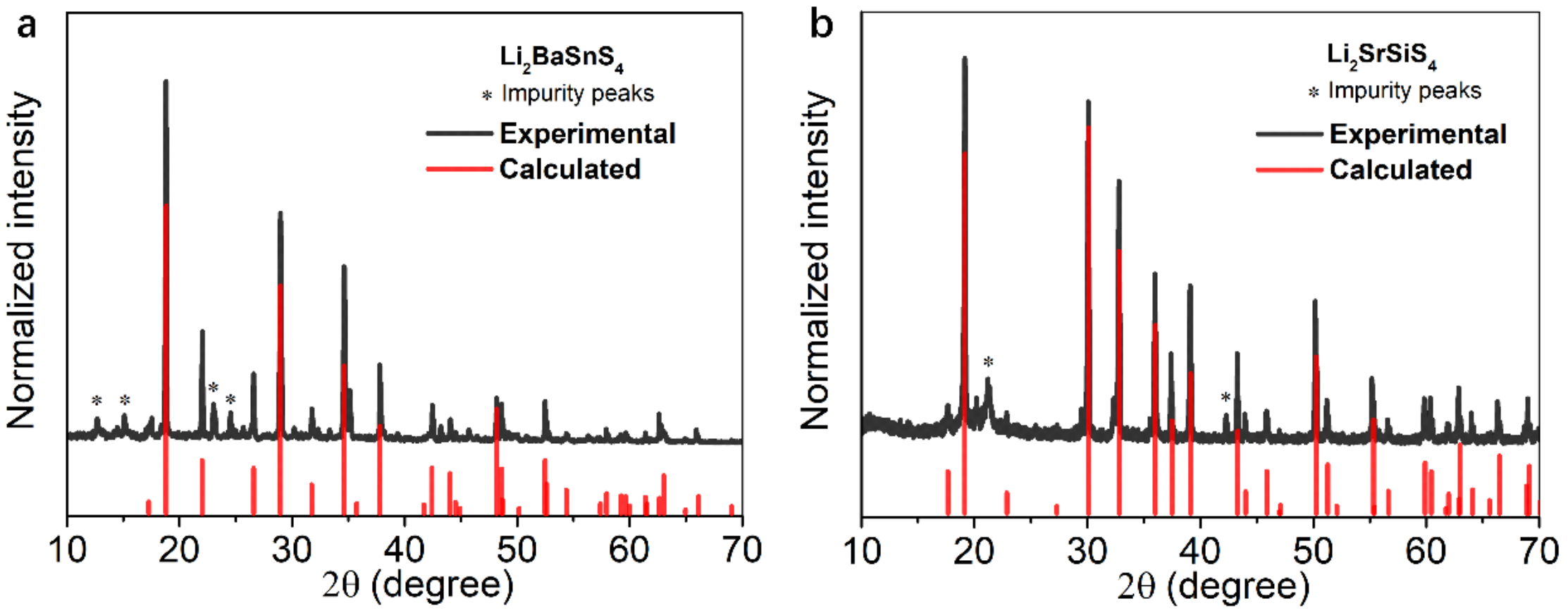

**Fig.S3** Powder X-ray diffraction (XRD) patterns of $Li_2BaSnS_4$ and $Li_2SrSiS_4$.

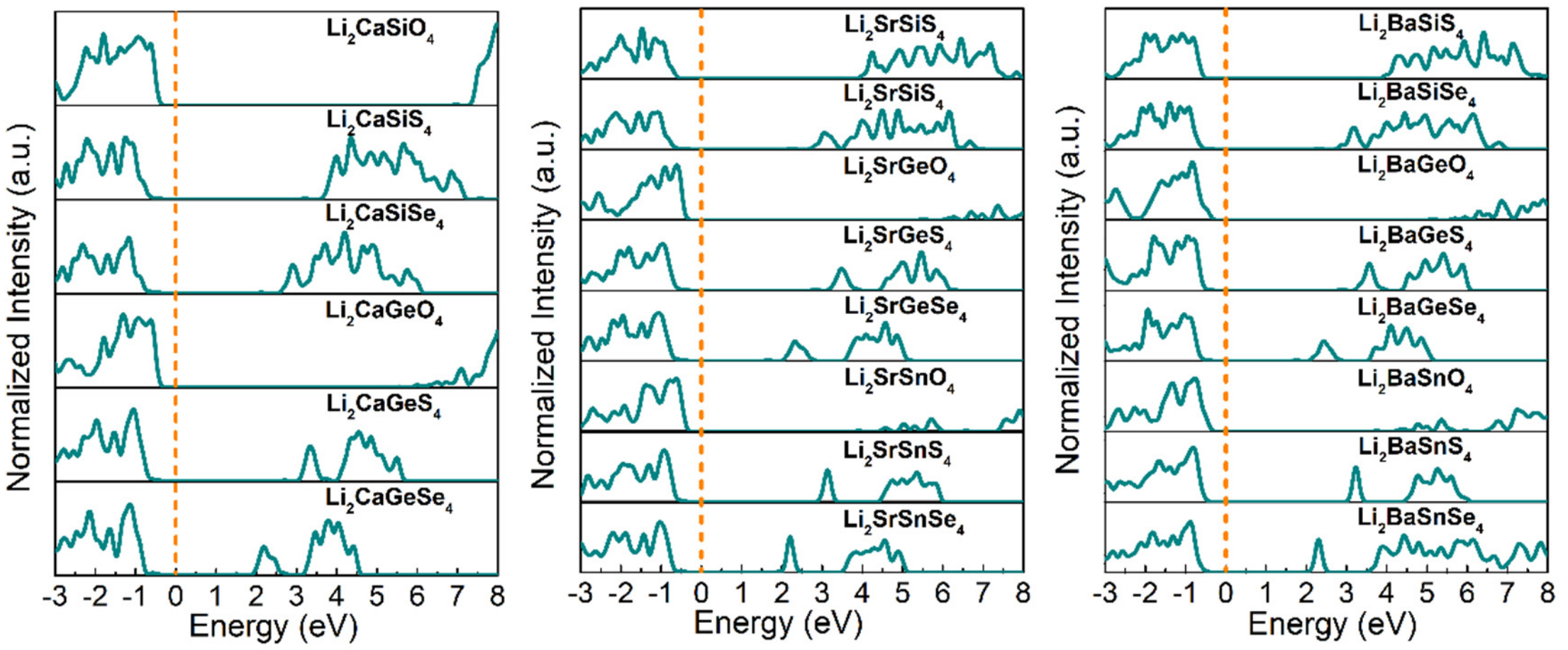

**Fig.S4** The calculated density of states by MBJ exchange potential for those compounds that have been synthesized experimentally or are thermodynamically and kinetically stable with I$\bar{4}$2m space group in $Li_2BMQ_4$ (B = Ca, Sr and Ba; M = Si, Ge and Sn and Q = O, S and Se) systems.

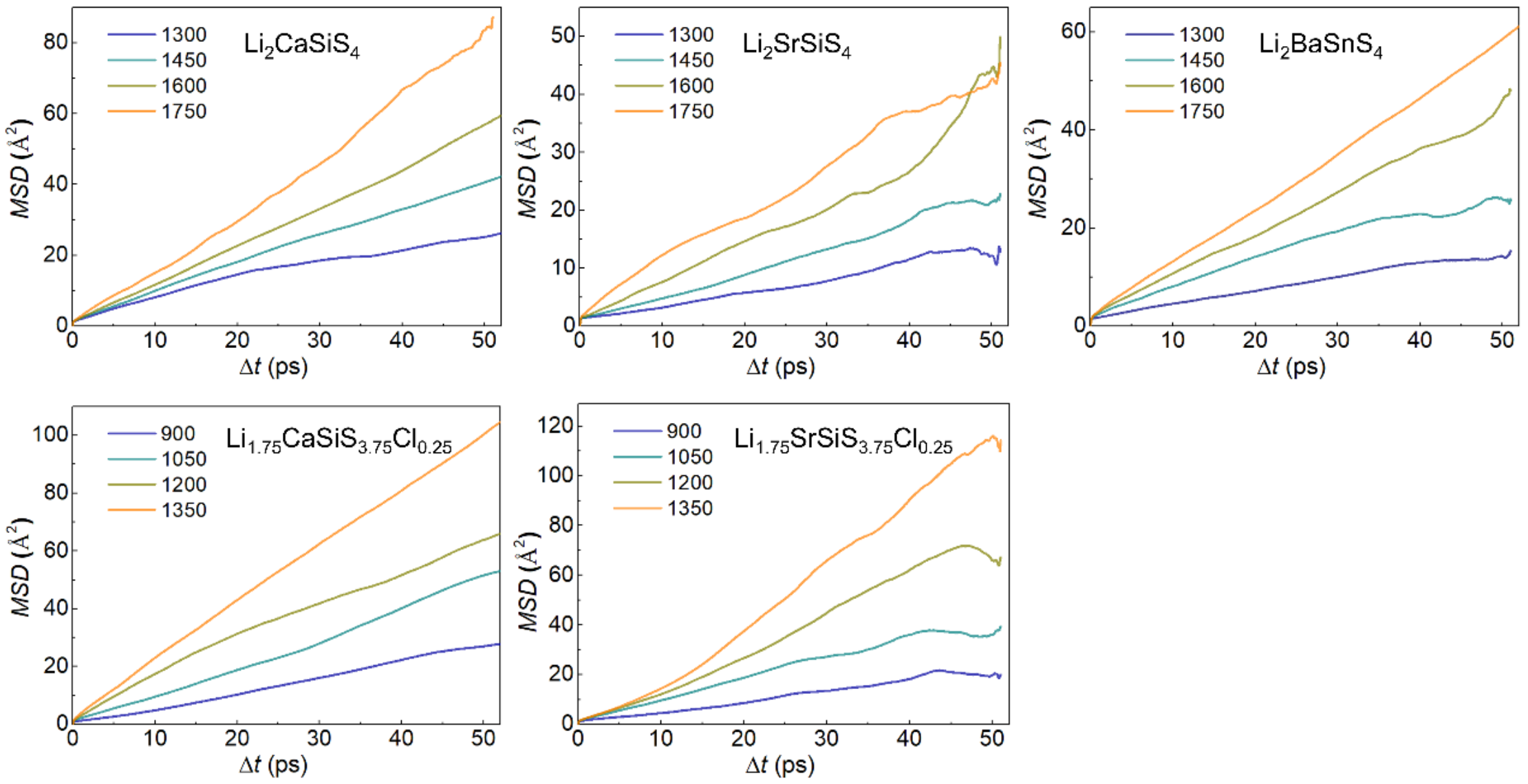

**Fig.S5** The *MSD*(mean-squared displacement)-Δ*t* curves from first-principles molecular dynamics (FPMD) simulations over 50 ps between 900 K and 1750 K for $Li_2CaSiS_4$, $Li_2SrSiS_4$, $Li_2BaSnS_4$, $Li_{1.75}CaSiS_{3.75}Cl_{0.25}$ and $Li_{1.75}SrSiS_{3.75}Cl_{0.25}$. The linear MSD-Δt dependence is utilized to apply the Einstein relation $D = \frac{\mathrm{MSD}(\Delta t)}{2d\Delta t} + D_{offset}$ for inferring the diffusivity, and the Arrhenius equation $D = D_0 \exp\left(-\frac{E_a}{kT}\right)$ is employed for linearly fitting log(D) against 1/T to derive the total activation energy

and ionic conductivity.

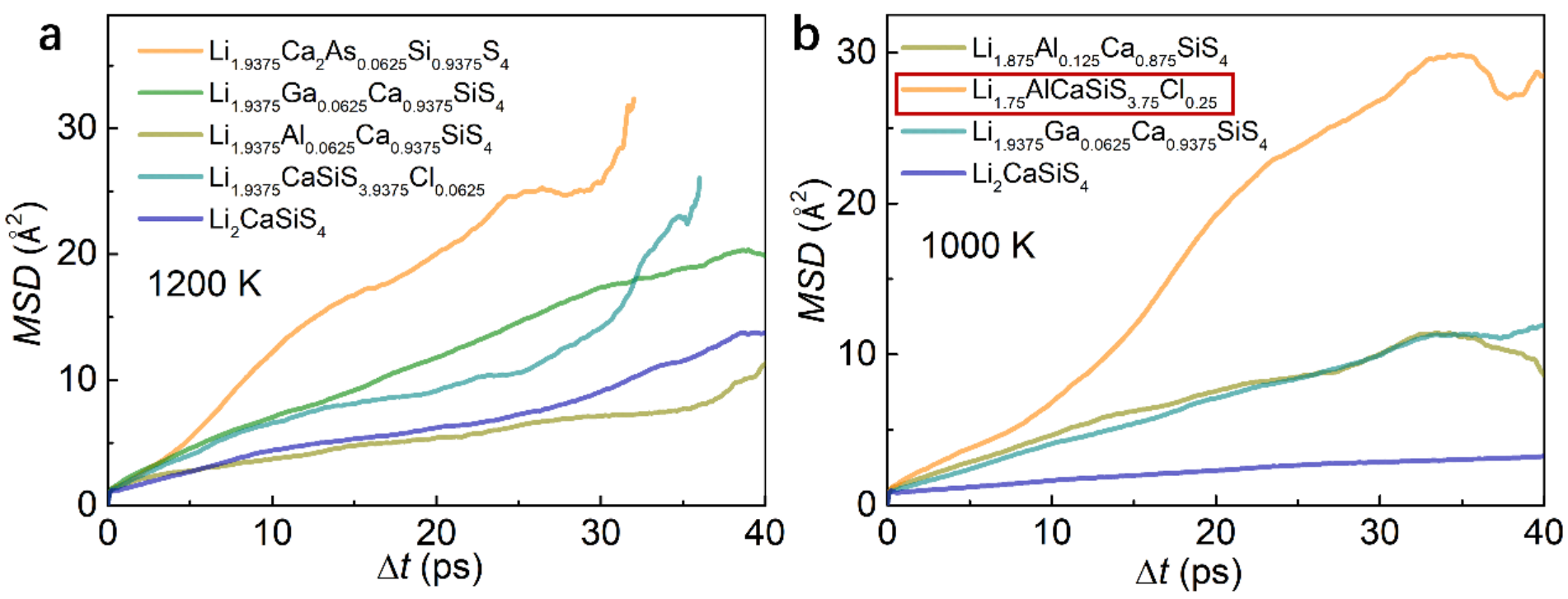

**Fig.S6** The MSD-Δt curves from first-principles molecular dynamics tests for $Li_2CaSiS_4$ and doped $Li_2CaSiS_4$ (with different doping elements such as Al, Ga, As and Cl at different concentrations) to choose appropriate doping element and doping concentrations. Mostly doped $Li_2CaSiS_4$ with some lithium vacancies exhibit larger MSD than the parent compound at the same temperature, indicating element doping could enhance $Li^+$ ionic conductivity. At lower doping concentrations, the enhancement of ionic conductivity is not significant, whereas higher doping concentrations may potentially enhance ionic conductivity better, albeit at the risk of inducing lattice instability. Taking these factors into account, we selected $Li_{1.75}CaSiS_{3.75}Cl_{0.25}$ configuration to perform FPMD with different temperatures aiming at estimating room-temperature ionic conductivity and activation energy, marked in the red box of **Fig.S6b**. The configurations of $Li_{1.75}SrSiS_{3.75}Cl_{0.25}$ is also constructed based on the configuration of $Li_{1.75}CaSiS_{3.75}Cl_{0.25}$.

## Reference

[1] Sun, J.-P.; McKeown Wessler, G. C.; Wang, T.; Zhu, T.; Blum, V.; Mitzi, D. B. Chem. Mater. **2020**, 32, 1636-1649.

[2] Kulshreshtha, C.; Shin, N.; Sohn, K.-S. Decay Behavior of $Li_2(Sr,Ba,Ca)SiO_4:Eu^{2+}$ Phosphors. *Electrochemical and Solid-State Letters* **2009**, *12*, J55-J57.

[3] Rodger, A. R.; Kuwano, J.; West, A. R. $Li^+$ ion conducting γ solid solutions in the systems $Li_4XO_4$-$Li_3YO_4$: X=Si, Ge, Ti; Y=P, As, V; $Li_4XO_4$-$LiZO_2$: Z=Al, Ga, Cr and $Li_4GeO_4$-$Li_2CaGeO_4$. *Solid State Ionics* **1985**, *15*, 185-198.

[4] 校凯，段炼，陈爽 可见光响应的光催化剂 $Li_2CaSnO_4$ 及其制备方法: 201610569389.0[P]. 2016-10-26.

[5] Yang, Y.; Wu, K.; Wu, X.; Zhang, B.; Gao, L. A new family of quaternary thiosilicates $SrA_2SiS_4$ (A = Li, Na, Cu) as promising infrared nonlinear optical crystals. *J. Mater. Chem. C* **2020**, *8*, 1762.

[6] Huang, S.; Li, G. Photoluminescence properties of $Li_2SrGeO_4:RE^{3+}$(RE=Ce/Tb/Dy) phosphors and enhanced luminescence through energy transfer between $Ce^{3+}$ and $Tb^{3+}/Dy^{3+}$. *Optical Materials* **2014**, *36*, 1555-1560.

[7] Wu, K.; Chu, Y.; Yang, Z.; Pan, S. $A_2SrM^{IV}S_4$ (A = Li, Na; $M^{IV}$ = Ge, Sn) concurrently exhibiting wide bandgaps and good nonlinear optical responses as new potential infrared nonlinear optical materials. *Chem. Sci.* **2019**, *10*, 3963.

[8] 李广卯，武奎，潘世烈 $Li_2BaSiSe_4$:一种具有显著二次倍频效应的新型复合金属硒化物，*科学通报*, *2019*, *64*, 1671-1678.

[9] Wu, K.; Zhang, B.; Yang, Z.; Pan, S. New Compressed Chalcopyrite-like $Li_2BaM^{IV}Q_4$ ($M^{IV}$ = Ge, Sn; Q = S, Se): Promising Infrared Nonlinear Optical Materials. *J. Am. Chem. Soc.* **2017**, *139,* 14885-14888.